\documentclass[lettersize,journal]{IEEEtran}
\usepackage[utf8]{inputenc}
\usepackage[T1]{fontenc}
\usepackage{graphicx,color,mathenv}
\usepackage[cmex10]{amsmath}
\usepackage{cite,amsthm,amssymb}
\usepackage{balance}
\usepackage[ruled,vlined]{algorithm2e}
\usepackage{amsthm}
\usepackage{subfigure,amsfonts}
\usepackage{float}
\usepackage{url}
\usepackage{xcolor}
\usepackage{tikz}
\usetikzlibrary{arrows.meta,calc}

\definecolor{rose}{RGB}{214,39,140}

\definecolor{orangecut}{RGB}{230,120,0}

\definecolor{twcgreen}{RGB}{0,140,70}
\newcommand\tw[1]{{\color{black}#1}}
\definecolor{dtcyan}{RGB}{0,125,165}

\usepackage{bbm}
\usepackage{cancel}
\usepackage{multirow}

\DeclareMathOperator*{\argmax}{arg\,max}

\DeclareMathAlphabet\mathbfcal{OMS}{cmsy}{b}{n}

\renewcommand{\Psi}{\varPsi}
\renewcommand{\Delta}{\varDelta}
\newtheorem{theorem}{Theorem}
\newtheorem{proposition}{Proposition}

\newtheorem{corollary}{Corollary}
\newtheorem{remark}{Remark}
\newtheorem{lemma}{Lemma}

\usepackage[all=normal,paragraphs=tight,floats=tight,mathspacing=tight,wordspacing=tight,charwidths=normal,mathdisplays=tight,leading=tight]{savetrees}

\usepackage{flushend}

\begin{document}

\title{\tw{One ISAC Array, Two Worst Cases: Hardware Impairments in Beamforming and AoA Estimation}}

\author{Xuan Chen, \IEEEmembership{Member, IEEE},  Luc Le Magoarou, \IEEEmembership{Member, IEEE}

\thanks{Xuan Chen and Luc Le Magoarou are with IETR, INSA Rennes, France (e-mail: firstname.lastname@insa-rennes.fr) }

\thanks{This work is supported by the French national research agency (MoBAIWL project, grant ANR-23-CE25-0013)}

}

\markboth{To be Submitted to IEEE Transactions on Wireless Communications}%
{Shell \MakeLowercase{\textit{et al.}}: A Sample Article Using IEEEtran.cls for IEEE Journals}


\maketitle

\begin{abstract}
In integrated sensing and communication (ISAC), the same base-station array beamforms toward users and estimates their angle of arrival (AoA). In this paper, we study how element-level gain, phase, and position errors affect both functions of a beam-scanning uniform linear array. Any array produces a correlation peak at the true angle, of finite height and width, surrounded by sidelobes. Impairments shift this peak, make it lower and wider, and raise the sidelobes relative to it. A single decomposition explains these three effects: the gain-weighted least-squares fit of the per-antenna phase errors splits them into a tilt, which shifts the peak, and a residual, whose power is taken from the peak and redistributed over the sidelobes according to its spatial spectrum. A beam steered toward the strongest direction suffers from the lower peak only, whereas the AoA estimate suffers from its shift even without noise, and also from the lowering and sidelobes with noise. The worst cases for the two ISAC functions are thus nearly disjoint, and we characterize both: the worst case for beamforming leaves the AoA estimate unbiased, and the worst case for localization loses little beamforming gain. The worst case for localization is a single defect at the junction between two subarrays, and its error, relative to the main-lobe width, does not decrease with the number of antennas. We also characterize random impairments and the effect of noise. In practical terms, our analysis shows that the calibration of a dual-function array must be assessed with respect to both functions.
\end{abstract}

\begin{IEEEkeywords}
AoA estimation, localization, integrated sensing and communication (ISAC), massive MIMO, array calibration, hardware impairments, worst-case analysis.
\end{IEEEkeywords}

\section{Introduction}
\label{sec:intro}

Multiple-input multiple-output (MIMO) arrays are a key enabler of fifth-generation (5G) and beyond networks \cite{marzetta2010noncooperative,larsson2014massive,massive_mimo_book}, and their spatial resolution also enables accurate localization and sensing \cite{10644093}, central to integrated sensing and communication (ISAC) \cite{liu2022isac,10888115,10478548,11162068} and intelligent transportation \cite{10584443}. In ISAC, the same base station (BS) array serves two functions: it beamforms toward the users, and it estimates their angle of arrival (AoA). Practical arrays, however, suffer from gain mismatches, phase offsets, and antenna position errors, due to manufacturing tolerances, calibration inaccuracies, aging, and environmental conditions \cite{7362822,9306107,10118823,11111976}. Moreover, large arrays are increasingly assembled from panels, each with its own radio-frequency chains and calibration \cite{9419749,6760595}, which makes these impairments correlated within each panel.

\begin{figure}[t]
\centering
\input{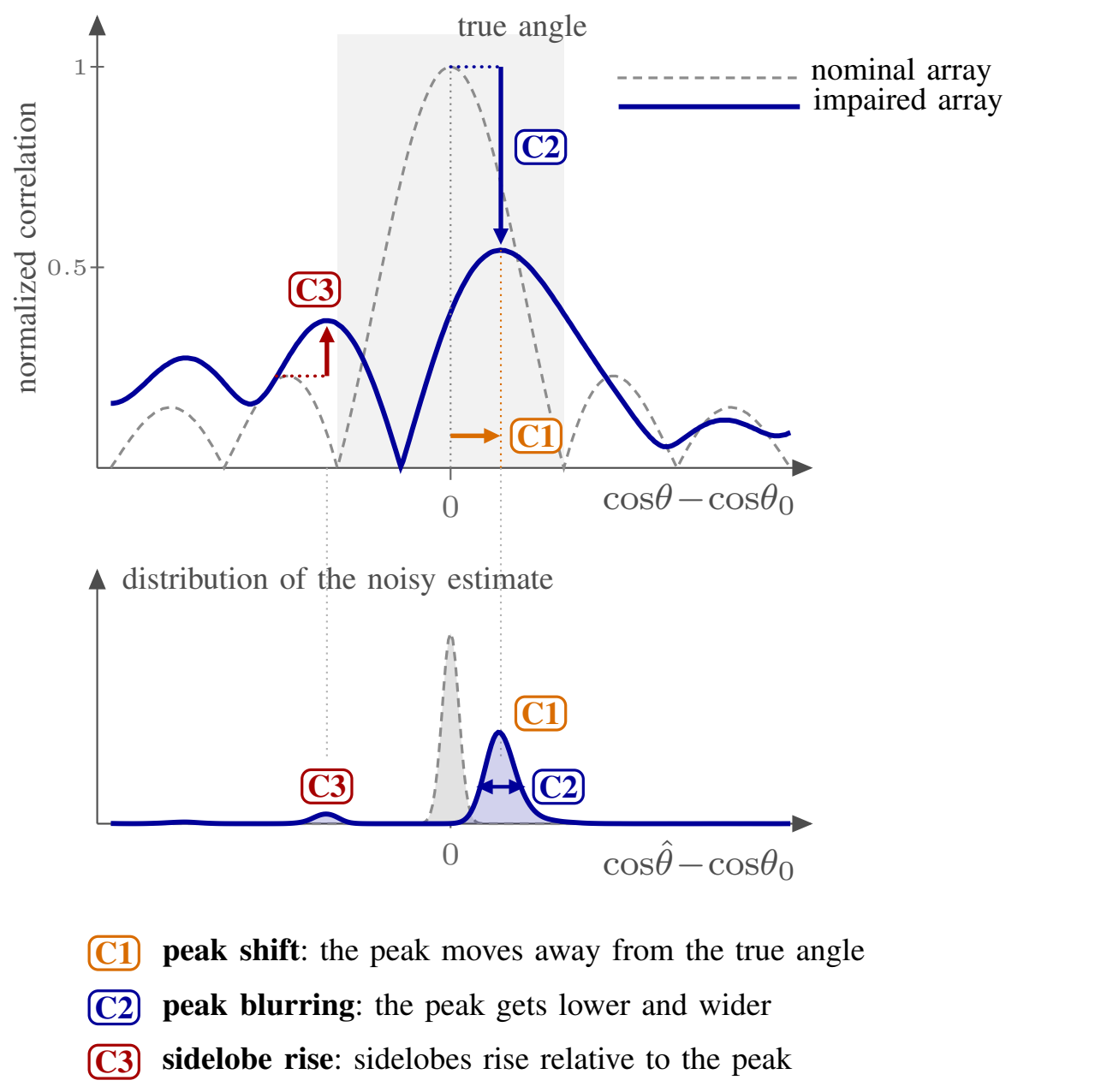}
\caption{Impact of array impairments on beam-scanning AoA estimation ($N=8$ antennas). Top: normalized correlation of a nominal (dashed) and an impaired (solid) array; the shaded area is the nominal main lobe. Bottom: distribution of the noisy AoA estimate over $40000$ noise draws, at an SNR of $8$~dB per antenna. The impairments shift the peak away from the true angle (C1, peak shift), make it lower and wider (C2, peak blurring), and raise the sidelobes relative to it (C3, sidelobe rise).}
\label{fig:teaser}
\end{figure}

When the BS scans its beam, any array produces a correlation peak at the true angle, of finite height and width, surrounded by sidelobes. As illustrated in Fig.~\ref{fig:teaser}, impairments have three consequences on this pattern:
\begin{itemize}
    \item[(C1)] \emph{peak shift}: the peak moves away from the true angle;
    \item[(C2)] \emph{peak blurring}: the peak gets lower and wider;
    \item[(C3)] \emph{sidelobe rise}: the sidelobes rise relative to the peak.
\end{itemize}
The two functions do not suffer from the same consequences. A beam steered toward the strongest direction only suffers from C2, whereas the AoA estimate suffers from C1 even without noise, and from C2 and C3 with noise. This raises an important question: do the impairments that are most harmful to one function also have the greatest impact on the other?

Most works on hardware impairments focus on system-level impairments, such as amplifier nonlinearities, phase noise, IQ imbalance, and quantization, modeled as additive distortions \cite{9306107,7342979} or memoryless nonlinearities \cite{9872081,massive_mimo_book,8891922}, and evaluate communication metrics. Array-level impairments directly perturb the array manifold: their influence on beamforming and calibration has been analyzed \cite{9772995,9172067,9146196}, their impact on localization mostly evaluated numerically \cite{10047354,10824218,11143375,10004742,guichemerre2026influencegainphasemismatches}, and learning-based methods have been proposed to compensate them for channel estimation \cite{yassine2022mpnet,klaimi2026mompnet}, AoA estimation \cite{chatelier2025diffmusic}, and ISAC \cite{10824218}, showing that impairments left by a communication-oriented calibration may still degrade sensing \cite{10824218}. Perturbation analyses have characterized subspace-based AoA estimators under small random sensor errors \cite{friedlander1990sensitivity,swindlehurst1992performance}, and antenna tolerance theory has quantified the effect of random element errors on the radiation pattern \cite{ruze1966antenna}: a linear phase error across the aperture steers the beam, whereas random phase errors lower the gain and raise the sidelobes. In our previous work \cite{11618421}, we characterized the worst-case beamforming loss caused by gain, phase, and position errors. However, no existing work explains C1 to C3 within a single framework, nor compares the worst cases of the two functions of an ISAC array.

In this paper, we formalize this classical intuition for a uniform linear array (ULA), and make the following contributions.
\begin{itemize}
    \item \emph{A decomposition explaining C1 to C3.} We split the phase errors into their gain-weighted least-squares fit across the aperture, the \emph{tilt}, and a \emph{residual}. The tilt exactly re-points the array (C1). The residual lowers and widens the peak (C2), and the energy it takes from the peak reappears in the sidelobes, at the angles given by its spatial spectrum (C3).
    \item \emph{A dispersion budget and two nearly disjoint worst cases.} The hardware bounds fix a budget on the dispersion of the phase errors, which can be spent on the tilt or on the residual, but not on both. Since beamforming toward the strongest direction only sees the residual and localization the tilt, their worst cases lie at the two ends of this budget: the worst case for beamforming does not bias the AoA estimate, and the worst case for localization, a phase step between two halves of the array with attenuated edges, loses little gain. Neither the beamforming gain nor the AoA accuracy alone is thus a sufficient calibration criterion. Random errors spend the budget at random: errors independent across antennas average out, whereas errors common to the antennas of a panel do not.
    \item \emph{The effect of noise.} The small-noise variance of the AoA estimate reduces to the Cram\'er--Rao bound for a nominal array and grows with C2, and C3 raises the signal-to-noise ratio (SNR) threshold below which the estimate jumps to a sidelobe. When the BS points its beam at the estimated angle, the bias costs nothing to communication, but the errors of the sensing step propagate to it.
\end{itemize}

The rest of the paper is organized as follows: Section~\ref{sec:system_model} presents the system model, Section~\ref{sec:theo} the decomposition, and Section~\ref{sec:effects} its effects C1 to C3. Section~\ref{sec:worst} studies the maximal effects of the impairments and their consequences for the two functions, Section~\ref{sec:noise} the effect of noise, and Section~\ref{sec:simu} provides numerical results.

\textbf{Notations}: Boldface letters $\boldsymbol{a}$ and normal font letters $a$ represent vectors and scalars, respectively; $\vec{\boldsymbol{a}}$ denotes a vector of the 3D Euclidean space. $\boldsymbol{a}^*$, $\boldsymbol{a}^T$ and $\boldsymbol{a}^H$ represent the complex conjugate, the transpose and the conjugate transpose of $\boldsymbol{a}$. $\mathbb{E}\{\cdot\}$ denotes the expectation and $\mathrm{sgn}(\cdot)$ the sign function. For any scalar-valued function $f$ of a single variable, $f'$ and $f''$ denote its first and second derivatives, respectively.

\section{System Model}
\label{sec:system_model}

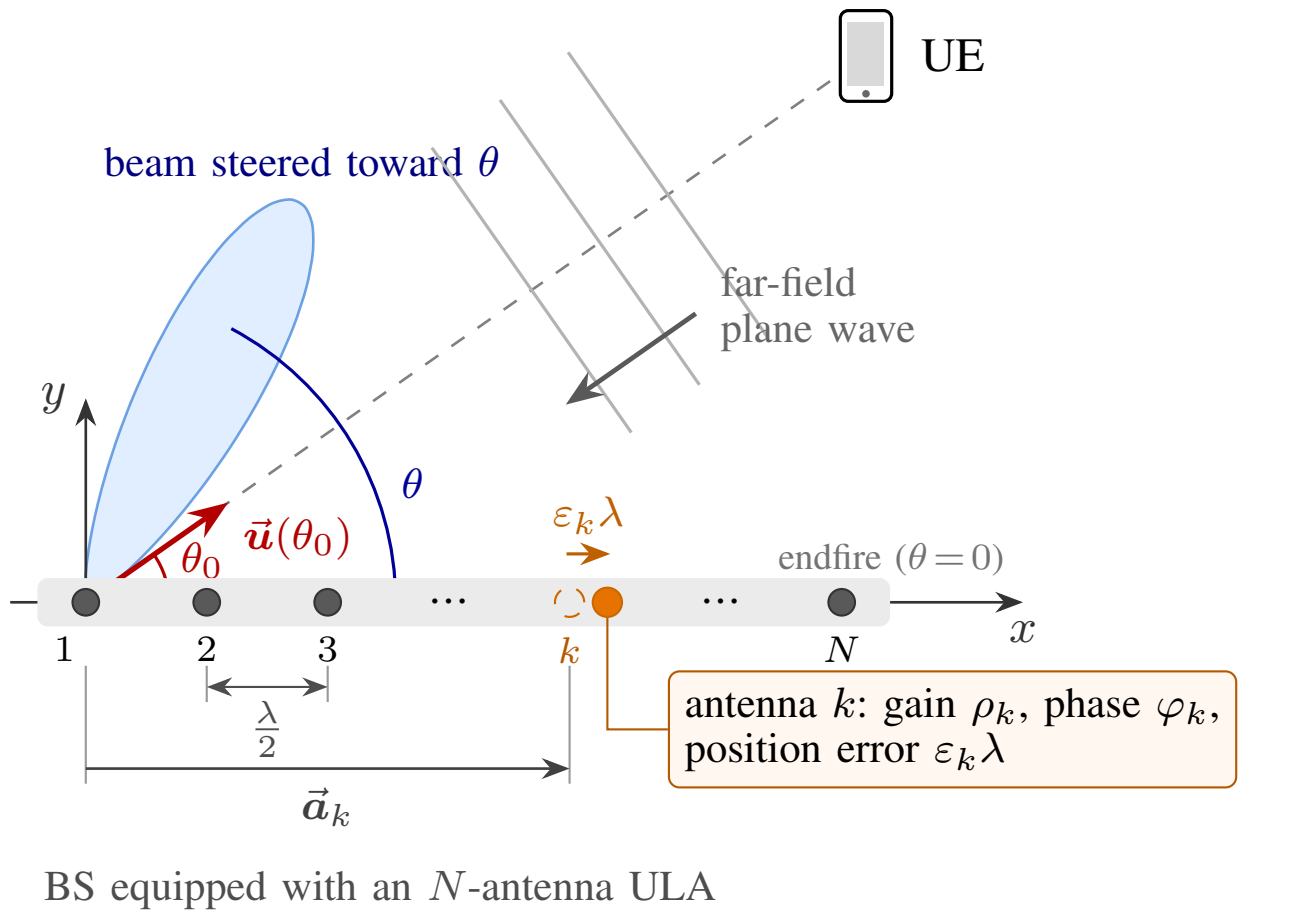
\begin{figure}[t]
\centering
\begin{tikzpicture}[
    >=Stealth, font=\small,
    ant/.style={circle, draw=black!80, fill=black!65, inner sep=0pt, minimum size=5pt},
    antk/.style={circle, draw=orange!80!black, fill=orange!90!black, inner sep=0pt, minimum size=5.5pt},
    antnom/.style={circle, draw=orange!80!black, dashed, inner sep=0pt, minimum size=5.5pt},
    dim/.style={<->, black!70, line width=0.4pt},
  ]
  \def\thz{35}      
  \def\ths{62}      
  \coordinate (O)  at (0,0);              
  \coordinate (UE) at (\thz:6.3);

  \fill[blue!55!cyan, opacity=0.12]
     plot[domain=-89:89, samples=90, smooth] ({\ths+\x}:{3.0*pow(cos(\x),22)}) -- cycle;
  \draw[blue!55!cyan!80!black, opacity=0.55, line width=0.5pt]
     plot[domain=-89:89, samples=90, smooth] ({\ths+\x}:{3.0*pow(cos(\x),22)});
  \node[blue!50!black, font=\footnotesize, anchor=south] at (\ths:3.05) {beam steered toward $\theta$};

  \draw[black!50, dashed, line width=0.5pt] (O) -- (UE);
  \foreach \t in {3.6,4.15,4.7}
     \draw[black!30, line width=0.6pt] ($(\thz:\t)+(\thz+90:1.15)$) -- ($(\thz:\t)+(\thz-90:1.15)$);
  \draw[->, line width=0.9pt, black!65] ($(\thz:4.4)+(\thz-90:0.75)$) -- ($(\thz:3.35)+(\thz-90:0.75)$);
  \node[black!60, font=\footnotesize, align=left, anchor=west] at ($(\thz:4.45)+(\thz-90:0.75)$) {far-field\\ plane wave};

  \begin{scope}[shift={(UE)}]
    \draw[rounded corners=1.5pt, fill=white, line width=0.6pt] (-0.17,-0.3) rectangle (0.17,0.3);
    \fill[black!15] (-0.12,-0.2) rectangle (0.12,0.22);
    \fill[black!60] (0,-0.25) circle (0.025);
    \node[right=2pt] at (0.17,0) {UE};
  \end{scope}

  \draw[->, black!80, line width=0.5pt] (-0.5,0) -- (6.2,0) node[below] {$x$};
  \draw[->, black!80, line width=0.5pt] (O) -- (0,1.35) node[left] {$y$};
  \node[black!55, font=\scriptsize, anchor=south east] at (6.2,0.05) {endfire ($\theta=0$)};

  \draw[->, line width=1pt, red!70!black] (O) -- (\thz:1.15) node[below right=-1pt, red!70!black, font=\footnotesize] {$\vec{\boldsymbol{u}}(\theta_0)$};
  \draw[red!70!black, line width=0.6pt] (0.55,0) arc[start angle=0, end angle=\thz, radius=0.55];
  \node[red!70!black, font=\footnotesize] at (20:0.82) {$\theta_0$};
  \draw[blue!60!black, line width=0.6pt] (2.05,0) arc[start angle=0, end angle=\ths, radius=2.05];
  \node[blue!60!black, font=\footnotesize] at (20:2.3) {$\theta$};

  \fill[black!8, rounded corners=2pt] (-0.32,-0.16) rectangle (5.32,0.16);
  \node[ant] (a1) at (0,0) {};
  \node[ant] (a2) at (0.8,0) {};
  \node[ant] (a3) at (1.6,0) {};
  \node at (2.4,0) {$\cdots$};
  \node[antnom] (aknom) at (3.2,0) {};
  \node[antk] (ak) at (3.45,0) {};
  \node at (4.2,0) {$\cdots$};
  \node[ant] (aN) at (5.0,0) {};
  \foreach \n/\lab in {a2/2, a3/3, aN/N}
     \node[below=3pt, font=\scriptsize] at (\n) {$\lab$};
  \node[below=3pt, xshift=-4pt, font=\scriptsize] at (a1) {$1$};
  \node[below=3pt, font=\scriptsize, orange!70!black] at (aknom) {$k$};

  \draw[black!40] ($(a2)+(0,-0.42)$) -- ($(a2)+(0,-0.66)$);
  \draw[black!40] ($(a3)+(0,-0.42)$) -- ($(a3)+(0,-0.66)$);
  \draw[dim] ($(a2)+(0,-0.56)$) -- node[below=-1pt, font=\footnotesize] {$\frac{\lambda}{2}$} ($(a3)+(0,-0.56)$);

  \draw[black!40] ($(aknom)+(0,-0.42)$) -- ($(aknom)+(0,-1.2)$);
  \draw[black!40] ($(a1)+(0,-0.42)$) -- ($(a1)+(0,-1.2)$);
  \draw[->, black!75, line width=0.5pt] (0,-1.1) -- node[below, font=\footnotesize] {$\vec{\boldsymbol{a}}_k$} ($(aknom)+(0,-1.1)$);

  \draw[->, orange!75!black, line width=0.6pt] ($(aknom)+(-0.02,0.32)$) -- ($(ak)+(0.02,0.32)$);
  \node[orange!75!black, font=\footnotesize, anchor=south] at ($(aknom)!0.5!(ak)+(0,0.33)$) {$\varepsilon_k\lambda$};

  \node[draw=orange!70!black, rounded corners=2pt, fill=orange!6, font=\footnotesize, align=left,
        anchor=north west, inner sep=3pt] (box) at (3.85,-0.45)
        {antenna $k$: gain $\rho_k$, phase $\varphi_k$,\\ position error $\varepsilon_k\lambda$};
  \draw[orange!70!black, line width=0.4pt] (ak.south) |- (box.west);

  \node[black!70, font=\footnotesize, anchor=north west] at (-0.4,-1.65) {BS equipped with an $N$-antenna ULA};
\end{tikzpicture}
\caption{System model: a BS equipped with an $N$-antenna ULA along the $x$-axis receives the signal of a far-field UE with AoA $\theta_0$ ($\theta=0$ corresponds to endfire) and scans its beam over $\theta$. The $k$th antenna is affected by a gain error $\rho_k$, a phase error $\varphi_k$, and a position error $\varepsilon_k\lambda$.}
\label{fig:system_model}
\end{figure}

\subsection{Signal Model and the Two Functions of the Array}
\label{sec:signal_model}

As shown in Fig.~\ref{fig:system_model}, we consider a BS equipped with a ULA of $N$ antennas spaced by $\frac{\lambda}{2}$, $\lambda$ being the wavelength, and a single-antenna user equipment (UE) located in the far field at the AoA $\theta_0$. The nominal position of the $k$th antenna, with respect to the first one, is $\vec{\boldsymbol{a}}_k=[\frac{\lambda}{2}(k-1),0,0]^T$, and the unit vector pointing toward the angle $\gamma$ is $\vec{\boldsymbol{u}}(\gamma)=[\cos(\gamma),\sin(\gamma),0]^T$, so that $\theta=0$ corresponds to the endfire direction and $\theta=90^\circ$ to the broadside one. The BS receives
\begin{equation}
\label{eq:y}
\boldsymbol{y}=\boldsymbol{h}(\theta_0)s+\boldsymbol{n},
\end{equation}
where $s$ is the transmitted symbol ($|s|=1$ without loss of generality), $\boldsymbol{h}(\theta_0)\in\mathbb{C}^{N}$ is the channel, and $\boldsymbol{n}\sim\mathcal{CN}(\boldsymbol{0},\sigma^2\boldsymbol{I}_N)$ is the noise. The BS only knows the nominal geometry of its array, through the steering vector
\begin{equation}
\label{eq:w}
\boldsymbol{w}(\theta)=\frac{1}{\sqrt{N}}\left[e^{-j\frac{2\pi}{\lambda}\vec{\boldsymbol{a}}_1\cdot\vec{\boldsymbol{u}}(\theta)},\ldots,e^{-j\frac{2\pi}{\lambda}\vec{\boldsymbol{a}}_N\cdot\vec{\boldsymbol{u}}(\theta)}\right]^T.
\end{equation}
It uses it for its two functions.
\begin{itemize}
    \item \emph{Localization.} The BS successively steers its beam toward candidate angles $\theta$, computes the correlation $z(\theta)=\boldsymbol{w}(\theta)^H\boldsymbol{y}$, and estimates the AoA as
    \begin{equation}
    \label{eq:est}
    \hat{\theta}=\argmax_{\theta}|z(\theta)|.
    \end{equation}
    \item \emph{Communication.} The BS beamforms with $\boldsymbol{w}(\theta_\mathrm{b})$, which yields a post-beamforming SNR proportional to $|\boldsymbol{w}(\theta_\mathrm{b})^H\boldsymbol{h}(\theta_0)|^2$. The beam direction $\theta_\mathrm{b}$ is either the true angle $\theta_0$, if it is known, or the direction that maximizes the received power, as in beam-sweeping procedures or when pointing at the estimated angle.
\end{itemize}
The analysis equally applies to the AoA estimation of a passive target from its echo, as long as the transmit illumination is not shaped by the impaired array (e.g., bistatic sensing); the monostatic case, where the impairments affect both transmission and reception, is left for future work.

\subsection{Hardware Impairment Model}
\label{sec:impairment_model}

With hardware impairments, the channel reads
\begin{equation}
\label{eq:h}
\boldsymbol{h}(\theta_0)=\beta\left[\tilde{g}_1e^{-j\frac{2\pi}{\lambda}\tilde{\vec{\boldsymbol{a}}}_1\cdot\vec{\boldsymbol{u}}(\theta_0)},\ldots,\tilde{g}_Ne^{-j\frac{2\pi}{\lambda}\tilde{\vec{\boldsymbol{a}}}_N\cdot\vec{\boldsymbol{u}}(\theta_0)}\right]^T,
\end{equation}
where $\beta$ is the channel attenuation, $\tilde{g}_k=\rho_ke^{j2\pi\varphi_k}$ is the complex gain of the $k$th antenna, and $\tilde{\vec{\boldsymbol{a}}}_k=\vec{\boldsymbol{a}}_k+\lambda[\varepsilon_k,0,0]^T$ its actual position. Three types of impairments are thus considered: \emph{gain errors} $\rho_k\in[1-\delta_g,1]$, with $\delta_g\in[0,1)$, which model element-to-element amplitude variations; \emph{phase errors} $\varphi_k\in[-\alpha_g,\alpha_g]$, expressed in cycles, which model different feed line lengths or impedance mismatches; and \emph{position errors} $\varepsilon_k\in[-\delta_p,\delta_p]$, expressed in wavelengths, which model mounting tolerances. The worst-case analysis of Section~\ref{sec:worst} only uses these bounds. The statistical analysis of Section~\ref{sec:stat} additionally assumes, as in \cite{10047354,11618421}, that the impairments are independent and uniformly distributed within their bounds.

Phase and position errors act alike: both add an extra phase to the contribution of each antenna. For the $k$th antenna, this extra phase is
\begin{equation}
\label{eq:Psi_k}
\Psi_k\triangleq2\pi\left(\varepsilon_k\cos(\theta_0)-\varphi_k\right),
\end{equation}
and its magnitude is at most $\Psi_{\max}\triangleq2\pi\big(\alpha_g+\delta_p|\cos(\theta_0)|\big)$, the maximal extra phase. Position errors are thus weighted by $|\mathrm{cos}(\theta_0)|$: they matter most at endfire and not at all at broadside, whereas phase errors matter equally in all directions. As in \cite{11618421}, we assume that $\alpha_g+\delta_p<\frac14$, which ensures $\Psi_{\max}<\frac{\pi}{2}$: all antennas still add constructively at the true angle ($\cos(\Psi_k)>0$).

\subsection{The Spatial Ambiguity Function}
\label{sec:ambiguity}

Using \eqref{eq:w} and \eqref{eq:h}, the response of the array to the UE, observed in the direction $\theta$, reads $\boldsymbol{w}(\theta)^H\boldsymbol{h}(\theta_0)=\frac{\beta}{\sqrt N}v(\Delta_0)$, with
\begin{equation}
\label{eq:v}
v(\Delta_0)\triangleq\sum_{k=1}^{N}\rho_k\,e^{j(2\pi d_k\Delta_0-\Psi_k)},\qquad \Delta_0\triangleq\cos(\theta)-\cos(\theta_0),
\end{equation}
where $d_k\triangleq\frac{k-1}{2}$ is the nominal position of the $k$th antenna in wavelengths. We call $|v(\Delta_0)|$, which is the correlation pattern obtained by beam scanning up to a normalization, the \emph{spatial ambiguity function} of the array: it governs both of its functions. It depends on $\theta$ only through the natural variable $\Delta_0$, and it is periodic of period $2$ in $\Delta_0$, the visible region $\theta\in[0,\pi]$, i.e., $\Delta_0\in[-1-\cos(\theta_0),1-\cos(\theta_0)]$, covering exactly one period. Without impairments ($\alpha_g=\delta_p=\delta_g=0$), it is simply the Dirichlet kernel
\begin{equation}
\label{eq:dirichlet}
|v(\Delta_0)|=\left|\frac{\sin(N\frac{\pi}{2}\Delta_0)}{\sin(\frac{\pi}{2}\Delta_0)}\right|,
\end{equation}
whose peak, of height $N$, is at $\Delta_0=0$, whose first nulls are at $\Delta_0=\pm\frac{2}{N}$, and whose highest sidelobes are about $13$~dB below the peak. We call $\frac{2}{N}$ the \emph{main-lobe half-width} and use it as the resolution unit of the array.

In the absence of noise, the two functions are characterized as follows.
\begin{itemize}
    \item \emph{Localization.} The AoA error is measured in the natural variable by
    \begin{equation}
    \label{eq:err}
    \hat{\Delta}_0\triangleq\cos(\hat\theta)-\cos(\theta_0)=\argmax_{\Delta_0}|v(\Delta_0)|,
    \end{equation}
    and the angle follows as $\hat\theta=\arccos(\cos(\theta_0)+\hat\Delta_0)\approx\theta_0-\hat\Delta_0/\sin(\theta_0)$ for small errors.
    \item \emph{Communication.} Relative to a nominal array perfectly pointed at the UE, the beamforming gain is $G_{\theta_0}\triangleq|v(0)|^2/N^2$ when the beam points at the true angle, and $G_\mathrm{peak}\triangleq\max_{\Delta_0}|v(\Delta_0)|^2/N^2$, the maximum being taken over the visible region, when it points at the strongest direction.
\end{itemize}
With noise, and taking $s=1$ without loss of generality since the noise is circularly symmetric, the correlation becomes $z(\theta)=\frac{\beta}{\sqrt N}v(\Delta_0)+\boldsymbol{w}(\theta)^H\boldsymbol{n}$, and the estimate becomes random; we denote it by $\tilde\Delta_0$, to distinguish it from the noiseless peak $\hat\Delta_0$. The SNR per antenna is $\mathrm{SNR}\triangleq|\beta|^2/\sigma^2$.

\section{Tilt and Residual Decomposition}
\label{sec:theo}

What are the respective effects of the different phase-error components? This section decomposes the phase errors into orthogonal components, which play completely separate roles in the ambiguity function, while Section~\ref{sec:effects} then derives the effect of each component.

\begin{figure}[t]
\centering
\includegraphics[width=\columnwidth]{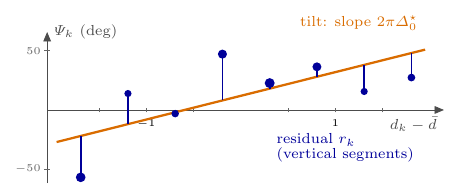}
\caption{Decomposition of the extra phases $\Psi_k$ of an impaired array ($N=8$ antennas, random impairments, marker size proportional to the gain $\rho_k$). The tilt is the gain-weighted least-squares line, of slope $2\pi\Delta_0^\star$; the residuals $r_k$ are the vertical deviations from this line.}
\label{fig:decomposition}
\end{figure}

Let $P\triangleq\sum_k\rho_k$, and for any sequences $(a_k)$ and $(b_k)$, let $\langle a,b\rangle_\rho\triangleq\sum_k\rho_ka_kb_k$ be their inner product weighted by the gains, $\bar a\triangleq\frac1P\langle a,1\rangle_\rho$ the gain-weighted mean of $(a_k)$, and $\mathrm{Var}_\rho(a)\triangleq\frac1P\langle a-\bar a,a-\bar a\rangle_\rho$ its gain-weighted variance over the antennas. In particular, $\bar d$ is the center of the array weighted by the gains, and $\mathrm{Var}_\rho(d)$ measures its effective aperture. With unit gains ($\rho_k=1$), these quantities reduce to the usual inner product, mean, and variance over the antennas. The reader may assume unit gains at first reading: as shown in Section~\ref{sec:tilt_effect}, gain errors create no effect by themselves, and only reweight the antennas. The constant sequence $(1)$ and the centered positions $(d_k-\bar d)$ are orthogonal for this inner product. Projecting the phase errors onto them gives
\begin{equation}
\label{eq:decomposition}
\Psi_k=\underbrace{\bar\Psi}_{\text{common phase}}+\underbrace{2\pi\Delta_0^\star(d_k-\bar d)}_{\text{tilt}}+\underbrace{r_k}_{\text{residual}},
\end{equation}
where the slope of the tilt is
\begin{equation}
\label{eq:tilt}
\Delta_0^\star\triangleq\frac{1}{2\pi}\,\frac{\langle d-\bar d,\Psi\rangle_\rho}{\langle d-\bar d,d-\bar d\rangle_\rho}=\frac{\sum_k\rho_k(d_k-\bar d)\Psi_k}{2\pi\sum_k\rho_k(d_k-\bar d)^2}.
\end{equation}
The \emph{tilt} is the component of the phase errors that varies linearly across the aperture, and the \emph{residual} $r_k$ is what remains. By construction, the residual is orthogonal to both other components, $\sum_k\rho_kr_k=0$ and $\sum_k\rho_k(d_k-\bar d)r_k=0$, so that the dispersion of the phase errors splits as
\begin{equation}
\label{eq:pythagoras}
\mathrm{Var}_\rho(\Psi)=\underbrace{(2\pi\Delta_0^\star)^2\,\mathrm{Var}_\rho(d)}_{\text{tilt}}+\underbrace{\mathrm{Var}_\rho(r)}_{\text{residual}}.
\end{equation}
Equivalently, the common phase and the tilt are the gain-weighted least-squares line fitted to the points $(d_k,\Psi_k)$, and the residuals are the deviations from this line, as illustrated in Fig.~\ref{fig:decomposition}. Note that $d_k$ are the nominal positions, known to the BS: position errors only enter through the extra phases $\Psi_k$, as in \eqref{eq:Psi_k}.

\section{Effects on the Ambiguity Function}
\label{sec:effects}

This section derives the effect of each component of the decomposition on the ambiguity function: the tilt in Section~\ref{sec:tilt_effect}, then the residual in Section~\ref{sec:residual_effect}. Section~\ref{sec:consequences} then summarizes which component generates each consequence C1 to C3 of Fig.~\ref{fig:teaser}.

\subsection{Effect of the Tilt: a Mere Re-Pointing}
\label{sec:tilt_effect}

Inserting \eqref{eq:decomposition} into \eqref{eq:v} directly gives the following exact factorization.

\begin{proposition}[Exact factorization]
\label{prop:factorization}
For any impairments,
\begin{equation}
\label{eq:factorization}
v(\Delta_0)=e^{j\chi}\,v_\mathrm{r}(\Delta_0-\Delta_0^\star),\qquad v_\mathrm{r}(x)\triangleq\sum_{k=1}^N\rho_k\,e^{j2\pi d_kx}\,e^{-jr_k},
\end{equation}
with $\chi\triangleq2\pi\Delta_0^\star\bar d-\bar\Psi$.
\end{proposition}

The three components of the phase errors thus play completely separate roles. The common phase only yields the global factor $e^{j\chi}$. The tilt only re-points the array: the ambiguity function is translated by $\Delta_0^\star$. The residual is the only component that changes the shape of the ambiguity function: $v_\mathrm{r}$ is the ambiguity function of an array whose phase errors reduce to the residual. In particular, when the residual vanishes, the impaired ambiguity function is exactly a translated copy of the nominal one, weighted by the gains.

\begin{corollary}[Pure tilt]
\label{cor:tilt}
If $r_k=0$ for all $k$, then $|v(\Delta_0)|=|V_\rho(\Delta_0-\Delta_0^\star)|$, where $V_\rho(x)\triangleq\sum_k\rho_ke^{j2\pi d_kx}$ is the gain-weighted nominal pattern. Since $|V_\rho(x)|\le P=V_\rho(0)$, the peak is exactly at $\Delta_0^\star$ (modulo the period $2$), with height $P$, whatever the size of the tilt.
\end{corollary}

In particular, gain errors alone ($\Psi_k=0$ for all $k$) create neither tilt nor residual: by Corollary~\ref{cor:tilt} with $\Delta_0^\star=0$, the peak stays exactly at the true angle, with the shape of the gain-weighted nominal pattern. In short, gain errors create no effect by themselves; they only reweight the antennas.

\subsection{Effect of the Residual}
\label{sec:residual_effect}

By Proposition~\ref{prop:factorization}, the effect of the residual is entirely described by $v_\mathrm{r}$. Expanding $e^{-jr_k}=1-jr_k-\frac{r_k^2}{2}+O(r_k^3)$ gives the \emph{second-order pattern}
\begin{equation}
\label{eq:v2}
v_2(x)\triangleq V_\rho(x)-j\,R(x)-\tfrac12R_2(x),
\end{equation}
where $R(x)\triangleq\sum_k\rho_kr_ke^{j2\pi d_kx}$ and $R_2(x)\triangleq\sum_k\rho_kr_k^2e^{j2\pi d_kx}$ are the spatial spectra of the gain-weighted residual and of its square. Geometrically, the residual rotates the phasor of each antenna by $r_k$. To first order, this rotation adds a component in quadrature, $-jr_k$, which does not change the length of the phasor; to second order, the in-phase term $-\frac{r_k^2}{2}$ shortens it, so that the phasor stays on the unit circle. By orthogonality of the decomposition, $R(0)=\sum_k\rho_kr_k=0$ and $R'(0)=j2\pi\sum_k\rho_k(d_k-\bar d)r_k=0$: at the tilt, the first-order term is zero and flat. This is also why the decomposition must be weighted by the gains: this weighting is precisely what cancels $R(0)$ and $R'(0)$. The following proposition describes the effect of the residual through the modulus of the second-order pattern, without any assumption on the size of the tilt.

\begin{proposition}[Effect of the residual]
\label{prop:residual}
For any gains and tilt, let $r_{\max}\triangleq\max_k|r_k|$ and $x\triangleq\Delta_0-\Delta_0^\star$. The second-order pattern approximates the ambiguity function within
\begin{equation}
\label{eq:v2bound}
\left|v(\Delta_0)-e^{j\chi}v_2(x)\right|\le\frac16\sum_k\rho_k|r_k|^3\le\frac{P}{6}\,r_{\max}\,\mathrm{Var}_\rho(r),
\end{equation}
and its modulus is, up to $O(r_{\max}^3)$,
\begin{multline}
\label{eq:mod2}
|v(\Delta_0)|^2\approx\underbrace{|V_\rho(x)|^2-\mathrm{Re}\big(V_\rho^*R_2\big)(x)}_{\text{lowered main lobe}}\\
+\underbrace{|R(x)|^2}_{\text{residual power}}+\underbrace{2\,\mathrm{Im}\big(V_\rho^*R\big)(x)}_{\text{interference}}.
\end{multline}
Consequently:
\begin{enumerate}
    \item[(i)] \emph{Peak location.} All the terms of \eqref{eq:mod2} have a zero slope at $x=0$: to second order in the residual, the peak is exactly at the tilt. The peak of the main lobe only moves at third order: as $r_{\max}\to0$, with $N$ and the gains fixed,
    \begin{equation}
    \label{eq:cubic}
    \hat\Delta_0=\Delta_0^\star-\frac{\sum_{k,l}\rho_k\rho_l(d_k-d_l)(r_k-r_l)^3}{12\pi\sum_{k,l}\rho_k\rho_l(d_k-d_l)^2}+O(r_{\max}^5),
    \end{equation}
    and $\hat\Delta_0=\Delta_0^\star$ exactly for pure tilts and for symmetric patterns;
    \item[(ii)] \emph{Main lobe.} Around the peak, \eqref{eq:mod2} reduces to the parabola $|v(\Delta_0)|^2/P^2\approx q(x)\triangleq1-\mathrm{Var}_\rho(r)-(2\pi)^2\,\mathrm{Var}_\rho(d)\,x^2$. More precisely, for all $\Delta_0$,
    \begin{equation}
    \label{eq:parabola}
    0\le\frac{|v(\Delta_0)|^2}{P^2}-q(x)\le\xi\triangleq\frac{1}{24P^2}\sum_{k,l}\rho_k\rho_l(\phi_k-\phi_l)^4,
    \end{equation}
    with $\phi_k\triangleq2\pi d_kx-r_k$: the parabola is an exact lower bound, accurate to fourth order in the phase differences $\phi_k-\phi_l$, i.e., for small residuals and for $|x|$ small compared with the inverse aperture;
    \item[(iii)] \emph{Energy balance.} Over the visible region, the means of the four terms of \eqref{eq:mod2} are $\sum_k\rho_k^2$, $-\sum_k\rho_k^2r_k^2$, $+\sum_k\rho_k^2r_k^2$, and $0$, respectively.
\end{enumerate}
\end{proposition}
\begin{proof}
See Appendix~\ref{app:residual}.
\end{proof}

The modulus \eqref{eq:mod2} reads as follows. The first term is the gain-weighted nominal pattern, re-pointed by the tilt, and lowered by the in-phase term, which is concentrated where $V_\rho$ is large, i.e., in the main lobe: at the peak, it removes $P^2\mathrm{Var}_\rho(r)$. The second term is the power spectrum of the residual, which adds in power to the nominal pattern and raises the sidelobes. The third term is an interference between the nominal pattern and the residual spectrum, in quadrature with each other: it has a zero slope at the peak and a zero mean, so that it only moves energy from one side of the peak to the other. By~(iii), the energy that the residual removes from the main lobe is exactly the energy it puts in the sidelobes. Statement~(i) holds as long as the main lobe remains the highest lobe; however, this condition is not always satisfied,  as discussed in  Section~\ref{sec:max_sidelobe}.

\subsection{What Generates C1, C2, and C3}
\label{sec:consequences}

The decomposition splits the dispersion of the phase errors into a tilt part and a residual part, by \eqref{eq:pythagoras}. Each consequence of Fig.~\ref{fig:teaser} is generated by one of them.

\emph{(C1) Peak shift: the tilt only.} By Corollary~\ref{cor:tilt} and Proposition~\ref{prop:residual}(i), the peak moves by the tilt $\Delta_0^\star$, exactly for a pure tilt and up to a third-order term in the residual otherwise. The shift is thus set by the tilt part $T\triangleq(2\pi\Delta_0^\star)^2\,\mathrm{Var}_\rho(d)$ of the dispersion: $|\Delta_0^\star|=\sqrt{T}/\big(2\pi\sqrt{\mathrm{Var}_\rho(d)}\big)$, so that a given tilt part shifts the peak all the more as the effective aperture is small. Without tilt, there is no peak shift: this is the case of gain errors alone, of a phase error common to all antennas, and of phase error patterns symmetric about the array center, for which $\Delta_0^\star=0$ and $|v|$ is even (see Appendix~\ref{app:residual}).

\emph{(C2) Peak blurring: the residual and the gains.} By Proposition~\ref{prop:residual}(ii), the peak has a height $|v(\Delta_0^\star)|\approx P\sqrt{1-\mathrm{Var}_\rho(r)}$, and at least this value, and $|v|^2$ has a curvature proportional to $P^2\,\mathrm{Var}_\rho(d)$ at the peak. The peak is thus lowered by the residual part of the dispersion and by the gain errors, through $P<N$; it is widened when the gains shrink the effective aperture $\mathrm{Var}_\rho(d)$, for instance when the edge antennas are attenuated. The tilt plays no role. At the true angle, by contrast, $|v(0)|=|v_\mathrm{r}(-\Delta_0^\star)|$, and the same argument as for \eqref{eq:parabola}, applied to the phases $\Psi_k$ themselves, leads to (see Appendix~\ref{app:residual})
\begin{equation}
\label{eq:v0}
0\le\frac{|v(0)|^2}{P^2}-\left[1-\mathrm{Var}_\rho(\Psi)\right]\le\frac{1}{24P^2}\sum_{k,l}\rho_k\rho_l(\Psi_k-\Psi_l)^4.
\end{equation}
The value at the true angle depends on the whole dispersion, tilt and residual parts together: it combines C1 and C2.

\emph{(C3) Sidelobe rise: the residual.} Where does the energy taken from the peak go? The following identity answers this question.

\begin{lemma}[Energy conservation]
\label{lem:energy}
For any impairments,
\begin{equation}
\label{eq:energy}
\frac{1}{2}\int_{-1-\cos(\theta_0)}^{1-\cos(\theta_0)}|v(\Delta_0)|^2\,d\Delta_0=\sum_{k=1}^N\rho_k^2.
\end{equation}
\end{lemma}
\begin{proof}
See Appendix~\ref{app:energy}.
\end{proof}

The average of $|v|^2$ over the visible region depends neither on the phase errors nor on the position errors: these impairments do not destroy the energy of the ambiguity function, they only redistribute it over the angles. The energy that the residual takes from the main lobe (C2) therefore reappears in the sidelobes (C3), through the power spectrum $|R|^2$ of the residual in \eqref{eq:mod2}, and Proposition~\ref{prop:residual}(iii) shows that the two exactly balance to second order: C2 and C3 are two faces of the same residual part of the dispersion. Where this energy goes depends on how the residual is arranged along the array: a white residual, as produced by independent errors, has a flat spectrum and raises a uniform floor of sidelobes, as described by antenna tolerance theory \cite{ruze1966antenna}; a periodic residual, as produced by panels with alternating offsets, concentrates it at its period, in a spurious lobe that can exceed the true peak (Section~\ref{sec:max_sidelobe}).

In short, the tilt part of the dispersion generates the peak shift only, the residual part generates both the peak blurring and the sidelobe rise, and the gains generate none of them by themselves, but lower the peak and reshape the effective aperture.

\section{Maximal Effects and the Two Functions of the Array}
\label{sec:worst}

Sections~\ref{sec:theo} and~\ref{sec:effects} show which component of the phase errors causes which consequence. We now ask how bounded hardware impairments can make each consequence as large as possible, and what this implies for the two functions of the array. The key constraint is a budget on the dispersion of the phase errors.

\subsection{A Dispersion Budget}
\label{sec:budget}

\begin{lemma}[Dispersion budget]
\label{lem:budget}
For any impairments satisfying the bounds of Section~\ref{sec:impairment_model},
\begin{equation}
\label{eq:budget}
(2\pi\Delta_0^\star)^2\,\mathrm{Var}_\rho(d)+\mathrm{Var}_\rho(r)=\mathrm{Var}_\rho(\Psi)\le\Psi_{\max}^2,
\end{equation}
with equality if and only if $|\Psi_k|=\Psi_{\max}$ for all $k$ and $\bar\Psi=0$.
\end{lemma}
\begin{proof}
The equality follows from \eqref{eq:pythagoras}. Moreover, $\mathrm{Var}_\rho(\Psi)=\frac1P\sum_k\rho_k\Psi_k^2-\bar\Psi^2\le\Psi_{\max}^2$, with equality if and only if $\Psi_k^2=\Psi_{\max}^2$ for all $k$ and $\bar\Psi=0$.
\end{proof}

The hardware bounds thus fix a budget on the dispersion of the phase errors, which can be spent on the tilt, causing a peak shift (C1), or on the residual, causing peak blurring (C2) and sidelobe rise (C3). The two cannot be maximal at the same time: an impairment configuration that is maximally harmful for one function necessarily spares the other. The gains do not enter the budget, but they act as a lever: they lower the peak through $P$, and they reshape the effective aperture $\mathrm{Var}_\rho(d)$.

\subsection{Maximal Peak Blurring: the Worst Case for Beamforming}
\label{sec:max_blur}

\begin{theorem}[Worst case for beamforming]
\label{theo:bf}
Let $N$ be even.
\begin{enumerate}
    \item[(i)] For any impairments, $G_{\theta_0}\ge(1-\delta_g)^2\cos^2\Psi_{\max}$, with equality if and only if all gains are minimal, i.e. $\rho_k=1-\delta_g$, and all extra phases reach their bounds with balanced signs: $\Psi_k=p_k\Psi_{\max}$ with $p_k\in\{-1,+1\}$ and $\sum_kp_k=0$, i.e., $\varepsilon_k=p_k\delta_p\,\mathrm{sgn}(\cos(\theta_0))$ and $\varphi_k=-p_k\alpha_g$.
    \item[(ii)] For any impairments, $G_\mathrm{peak}\ge(1-\delta_g)^2\cos^2\Psi_{\max}$, with equality if and only if the configuration satisfies the conditions of~(i) and its highest peak lies at the true angle. Such configurations do not shift the peak: $\hat\Delta_0=0$.
\end{enumerate}
\end{theorem}
\begin{proof}
See Appendix~\ref{app:bf}.
\end{proof}

We call \emph{pattern} the sequence of signs $(p_1,\ldots,p_N)$ of the extra phases in Theorem~\ref{theo:bf}(i): for instance, the split pattern $(+,\ldots,+,-,\ldots,-)$, the alternating pattern $(+,-,+,-,\ldots)$, or the symmetric patterns, which are palindromic, such as $(-,+,+,-)$. Statement~(i) recovers the result of \cite{11618421}: the beamforming loss toward the true angle only depends on how many signs are positive or negative, not on their arrangement, so that all the $\binom{N}{N/2}$ balanced patterns are worst cases. In terms of the budget of Lemma~\ref{lem:budget}, they all spend the whole budget, with minimal gains, but they split it very differently between tilt and residual. This split does not matter for $G_{\theta_0}$, which depends on the total dispersion, but it is crucial for the other function. For instance, with $N=4$ and $\Psi_k=p_k\Psi_{\max}$, the nominal antenna positions relative to the array center are $d_k-\bar d=(-\frac34,-\frac14,\frac14,\frac34)$, in wavelengths, and \eqref{eq:tilt} gives $\Delta_0^\star=\frac{2\Psi_{\max}}{5\pi}\sum_k(d_k-\bar d)p_k$: the split pattern $(-,-,+,+)$ gives $\Delta_0^\star=\frac{4}{5\pi}\Psi_{\max}\approx0.25\,\Psi_{\max}$, the alternating pattern $(-,+,-,+)$ gives half this value, and the symmetric pattern $(-,+,+,-)$ gives $0$, although all three are worst cases for beamforming toward the true angle.

Statement~(ii) shows that, when the beam points at the strongest direction, the worst loss is the same, but it is only reached by the configurations that spend the whole budget on the residual, without any peak shift. For $N$ a multiple of $4$, the symmetric patterns, e.g., $(-,+,+,-)$ repeated, reach it as long as their highest peak stays at the true angle. Their period-$4$ structure creates replicas at $\Delta_0=\pm\frac12$, of height close to $\frac{\sqrt2}{2}N(1-\delta_g)\sin\Psi_{\max}$, which remain below the true peak $N(1-\delta_g)\cos\Psi_{\max}$ for $\Psi_{\max}<\arctan\sqrt2\approx54.7^\circ$, as we verified numerically for $N$ up to $64$. The worst case for beamforming toward the strongest direction is thus a maximal peak blurring without any peak shift.

\subsection{Maximal Peak Shift: the Worst Case for Localization}
\label{sec:max_shift}

By \eqref{eq:tilt} and $|\Psi_k|\le\Psi_{\max}$, the tilt of any configuration satisfies
\begin{equation}
\label{eq:tilt_bound}
|\Delta_0^\star|\le\frac{\Psi_{\max}}{2\pi}\,Q(\boldsymbol\rho),\qquad Q(\boldsymbol\rho)\triangleq\frac{\sum_k\rho_k|d_k-\bar d|}{\sum_k\rho_k(d_k-\bar d)^2},
\end{equation}
with equality if and only if $\Psi_k=\pm\Psi_{\max}\,\mathrm{sgn}(d_k-\bar d)$, with the same sign for all $k$: a phase step at the center, which spends the whole budget toward the tilt. The quantity $Q(\boldsymbol\rho)$ measures how efficiently a gain profile converts a phase step into a tilt. Let $c\triangleq\frac{N-1}{4}$ be the geometric center of the array; for gain profiles symmetric about it, $\bar d=c$. Let $Q^\star$ denote the maximum of $Q$ over such profiles. Then the worst case of localization can be summarized as follows:

\begin{theorem}[Worst case for localization]
\label{theo:loc}
Among gain profiles symmetric about the array center, the largest tilt is $\frac{\Psi_{\max}}{2\pi}Q^\star$. It is reached exactly by the two mirror configurations indexed by $\varsigma\in\{-1,+1\}$:
\begin{itemize}
    \item $\varepsilon_k=-\varsigma\,\delta_p\,\mathrm{sgn}(\cos(\theta_0))\,\mathrm{sgn}(d_k-c)$;
    \item $\varphi_k=\varsigma\,\alpha_g\,\mathrm{sgn}(d_k-c)$;
    \item $\rho_k=1$ if $|d_k-c|<\frac{1}{Q^\star}$, and $\rho_k=1-\delta_g$ otherwise,
\end{itemize}
which yield $\Delta_0^\star=-\varsigma\,\frac{\Psi_{\max}}{2\pi}Q^\star$. Moreover, in the large-$N$ regime of Corollary~\ref{cor:Qstar}, no gain profile, symmetric or not, yields a larger tilt.
\end{theorem}
\begin{proof}
See Appendix~\ref{app:loc}.
\end{proof}

By Proposition~\ref{prop:residual}(i), the peak of these configurations is at their tilt, up to third order in their residual. The restriction to symmetric gain profiles is only needed for the exact statement at finite $N$, where it does not appear to be binding either: an exhaustive search over all $2^N$ vertex gain profiles for $N\le16$, and a continuous optimization over the whole box $[1-\delta_g,1]^N$ for $N$ up to $64$, always returned the symmetric profile of Theorem~\ref{theo:loc}.

The structure of this worst case can be easily understood. For $\varsigma=+1$ and $\cos(\theta_0)>0$, the two halves of the array are shifted by $\delta_p$ toward the center, which shrinks the array; for $\varsigma=-1$, the array is stretched. The phase errors add an opposite offset to each half, with the same sign as the one induced by the displacement, so that $\Psi_k$ reaches its bound. Finally, the central antennas keep their nominal gain, whereas those beyond the distance $\frac{1}{Q^\star}$ from the center are attenuated: this shrinks the effective aperture $\mathrm{Var}_\rho(d)$, so that the same phase step produces a steeper tilt. The gains do not create the peak shift, but they amplify it.

The maximum $Q^\star$ is the fixed point of the threshold rule of Theorem~\ref{theo:loc}, and has no closed form at finite $N$. Corollary~\ref{cor:Qstar} gives it explicitly in the large-$N$ regime, together with the size of the unit-gain center of the array. Theorem~\ref{theo:locerr} then turns it into the worst-case AoA error in closed form, which makes its dependence on $N$, $\Psi_{\max}$, and $\delta_g$ explicit.

\begin{corollary}
\label{cor:Qstar}
In the large-$N$ regime, $Q^\star\approx\frac{4}{N\tau}$, and the number of antennas with unit gain on each side of the center is $M_\mathrm{uni}\approx\frac{\tau N}{2}$, where $\tau$ is the unique real root in $(0,\frac23]$ of $\delta_g\tau^3+3(1-\delta_g)\tau-2(1-\delta_g)=0$, equal to $\frac23$ for $\delta_g=0$ and given for $\delta_g>0$ by
\begin{equation}
\label{eq:tau}
\tau=\sqrt[3]{\frac{1-\delta_g}{\delta_g}\left(1+\frac{1}{\sqrt{\delta_g}}\right)}-\sqrt[3]{\frac{1-\delta_g}{\delta_g}\left(\frac{1}{\sqrt{\delta_g}}-1\right)}.
\end{equation}
\end{corollary}
\begin{proof}
See Appendix~\ref{app:Qstar}.
\end{proof}

\begin{theorem}[Worst-case AoA error]
\label{theo:locerr}
The worst-case peak shift of Theorem~\ref{theo:loc}, $\Delta_0^\mathrm{wst}\triangleq\frac{\Psi_{\max}}{2\pi}Q^\star$, satisfies
\begin{equation}
\label{eq:wst}
\Delta_0^\mathrm{wst}\approx\frac{3\Psi_{\max}}{\pi N}\,\Lambda(\delta_g),\qquad \Lambda(\delta_g)\triangleq\frac{2}{3\tau}.
\end{equation}
In the angular domain, the estimate is $\hat\theta=\arccos\big(\cos(\theta_0)-\varsigma\Delta_0^\mathrm{wst}\big)$. For $0<\theta_0\le90^\circ$, the configuration $\varsigma=-1$ yields the larger angular error, provided that $\cos(\theta_0)+\Delta_0^\mathrm{wst}<1$, i.e., $N>\frac{3\Psi_{\max}\Lambda(\delta_g)}{\pi(1-\cos(\theta_0))}$.
\end{theorem}
\begin{proof}
See Appendix~\ref{app:Qstar}.
\end{proof}

The lever factor $\Lambda(\delta_g)\ge1$ quantifies the lever effect of the gains: it is the ratio $Q^\star/Q_\mathrm{unif}$ between the largest tilt per unit phase step and its value $Q_\mathrm{unif}\approx\frac6N$ with uniform gains. Attenuating the edge antennas reduces the moment of inertia $\sum_k\rho_k(d_k-\bar d)^2$ of the least-squares fit more than its lever arm $\sum_k\rho_k|d_k-\bar d|$, which steepens the fitted slope. It increases with $\delta_g$: $\Lambda(0)=1$, $\Lambda(0.1)\approx1.02$, $\Lambda(0.3)\approx1.06$, and $\Lambda(0.7)\approx1.23$. The same attenuation also widens the main lobe, whose curvature is proportional to $\mathrm{Var}_\rho(d)$: for $\delta_g=0.7$, the effective aperture of this configuration is about $60\%$ of the uniform one, which widens the main lobe by a factor of about $1.29$. Relative to the nominal main-lobe width, i.e., to the resolution expected from the array, the peak shift is thus amplified by $\Lambda(\delta_g)$; relative to the actual main lobe of the attenuated array, it is not. 

In absolute terms, the worst-case error decreases as $\frac1N$, but the main lobe shrinks at the same rate: relative to the main-lobe half-width $\frac2N$, the worst-case peak shift is $\frac{3}{2\pi}\Psi_{\max}\Lambda(\delta_g)$, which does not depend on $N$. In short, larger arrays are more accurate, but not more tolerant to hardware impairments.

\subsection{Maximal Sidelobe Rise: Validity of the Main-Lobe Analysis}
\label{sec:max_sidelobe}

Theorem~\ref{theo:loc} maximizes the shift of the main-lobe peak. Larger AoA errors arise when a sidelobe exceeds the main lobe, i.e., for a maximal sidelobe rise, which a periodic residual produces most efficiently (Section~\ref{sec:consequences}). We do not analyze this case theoretically; its consequences with noise are illustrated numerically in Section~\ref{sec:simu}. We only note that it delimits the validity of Theorem~\ref{theo:loc}. For instance, with unit gains and phase errors alternating between $\pm\Psi_{\max}$ from one antenna to the next, the ambiguity function has a replica of the main lobe at $\Delta_0=\pm1$, of height $N\sin\Psi_{\max}$, against $N\cos\Psi_{\max}$ for the true peak: it exceeds the true peak as soon as $\Psi_{\max}>45^\circ$. With strong gain errors, gross errors appear numerically from $\Psi_{\max}\approx38^\circ$ for $\delta_g=0.7$. Below these thresholds, Theorem~\ref{theo:loc} gives the worst AoA error.

\subsection{Consequences for the Two Functions}
\label{sec:two_functions}

\begin{table}[t]
\caption{Consequences of the impairments affecting each function: C1 peak shift, C2 peak blurring, C3 sidelobe rise.}
\label{tab:who}
\centering
\footnotesize
\renewcommand{\arraystretch}{1.2}
\begin{tabular}{lccc}
\hline
& C1 & C2 & C3 \\
\hline
Beamforming toward $\theta_0$ ($G_{\theta_0}$) & \checkmark & \checkmark & -- \\
Beamforming toward the peak ($G_\mathrm{peak}$) & -- & \checkmark & -- \\
AoA estimation, without noise & \checkmark & -- & -- \\
AoA estimation, with noise & \checkmark & \checkmark & \checkmark \\
\hline
\end{tabular}
\end{table}

Table~\ref{tab:who} summarizes which consequences affect each function, following Section~\ref{sec:consequences}: the peak height only depends on the residual and the gains, whereas the value at the true angle combines the peak shift and the peak blurring. Proposition~\ref{prop:factorization} makes the former exact: since $v$ is periodic of period $2$ and the visible region covers exactly one period, then
\begin{equation}
\label{eq:gpeak}
G_\mathrm{peak}=\frac{\max_x|v_\mathrm{r}(x)|^2}{N^2}
\end{equation}
does not depend on the tilt at all: beamforming toward the strongest direction is exactly blind to it. Without noise, the AoA estimate only suffers from the peak shift; with noise, it also suffers from the peak blurring and the sidelobe rise (Section~\ref{sec:noise}), and a beam pointed at the estimated angle inherits these errors.

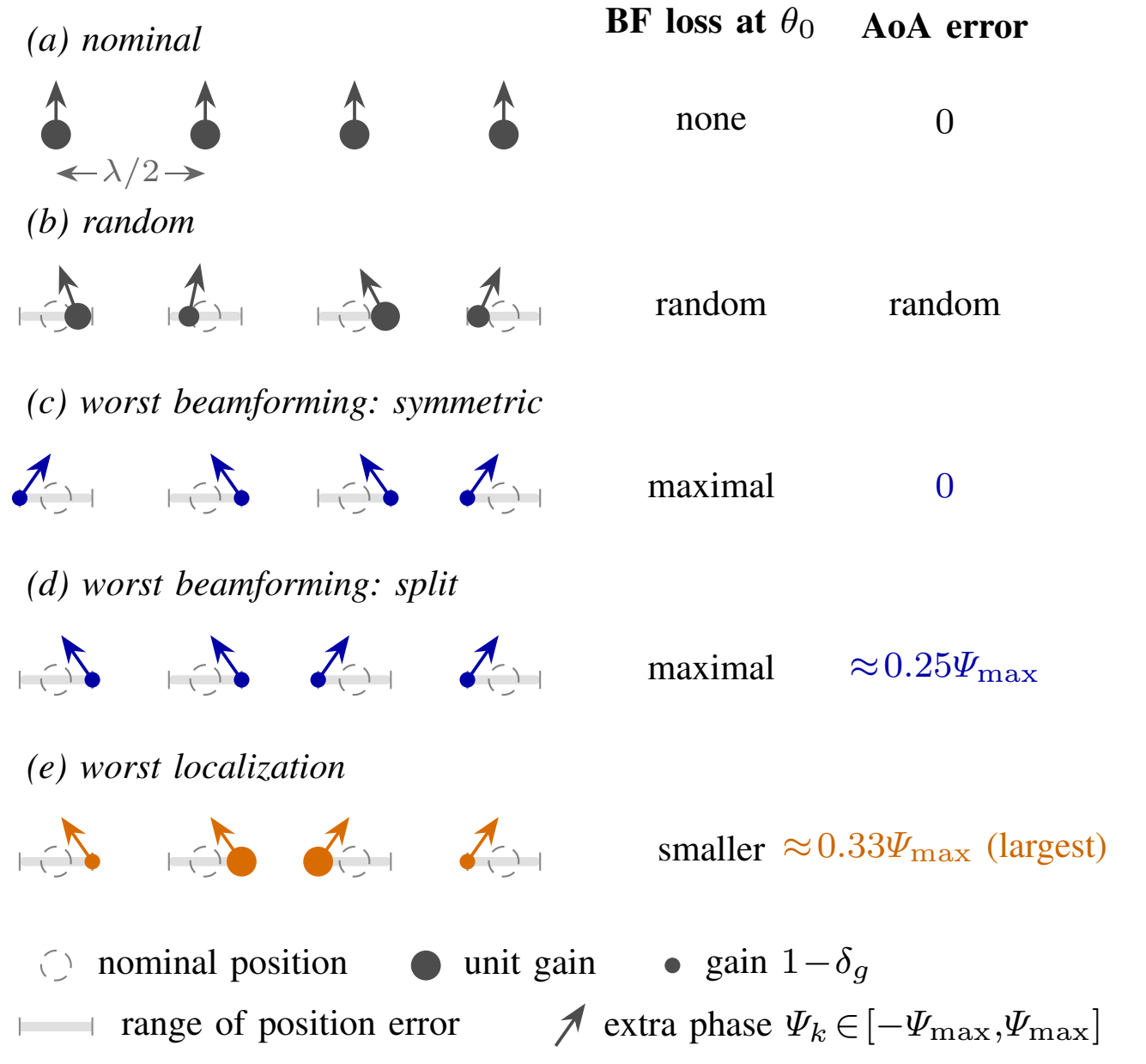
\begin{figure}[t]
\centering
\begin{tikzpicture}[>=Stealth, font=\footnotesize]
  \def\spc{1.15}   
  \def\dpd{0.28}   
  \def\Pmax{35}    
  \def\rmax{0.115} 
  \def\rmin{0.06}  
  \def\colA{5.05} \def\colB{6.85}
  \newcommand{\ant}[6]{%
    \pgfmathsetmacro{\xn}{(#1-1)*\spc}
    \pgfmathsetmacro{\xa}{\xn+#3*\dpd}
    \fill[black!12, rounded corners=1pt] (\xn-\dpd,#2-0.035) rectangle (\xn+\dpd,#2+0.035);
    \draw[black!45] (\xn-\dpd,#2-0.07) -- (\xn-\dpd,#2+0.07) (\xn+\dpd,#2-0.07) -- (\xn+\dpd,#2+0.07);
    \draw[black!50, densely dashed] (\xn,#2) circle (\rmax);
    \draw[->, line width=0.7pt, #6] (\xa,#2) -- ++(90+#5:0.42);
    \fill[#6] (\xa,#2) circle (#4);
  }
  \newcommand{\antnominal}[2]{%
    \pgfmathsetmacro{\xn}{(#1-1)*\spc}
    \draw[->, line width=0.7pt, black!70] (\xn,#2) -- ++(90:0.42);
    \fill[black!70] (\xn,#2) circle (\rmax);
  }
  \node[font=\footnotesize\bfseries, anchor=south] at (\colA,0.62) {BF loss at $\theta_0$};
  \node[font=\footnotesize\bfseries, anchor=south] at (\colB,0.62) {AoA error};

  \def\y{0}
  \node[anchor=south west, font=\footnotesize\itshape] at (-0.35,\y+0.47) {(a) nominal};
  \foreach \i in {1,...,4} {\antnominal{\i}{\y}}
  \node at (\colA,\y+0.1) {none}; \node at (\colB,\y+0.1) {$0$};
  \draw[<->, black!60, line width=0.4pt] (0,\y-0.3) -- node[midway, fill=white, inner sep=1pt, font=\scriptsize] {$\lambda/2$} (\spc,\y-0.3);

  \def\y{-1.4}
  \node[anchor=south west, font=\footnotesize\itshape] at (-0.35,\y+0.47) {(b) random};
  \ant{1}{\y}{0.6}{0.103}{20}{black!70}
  \ant{2}{\y}{-0.45}{0.08}{-12}{black!70}
  \ant{3}{\y}{0.85}{0.112}{28}{black!70}
  \ant{4}{\y}{-0.7}{0.087}{-24}{black!70}
  \node at (\colA,\y+0.1) {random}; \node at (\colB,\y+0.1) {random};

  \def\y{-2.8}
  \node[anchor=south west, font=\footnotesize\itshape] at (-0.35,\y+0.47) {(c) worst beamforming: symmetric};
  \ant{1}{\y}{-1}{\rmin}{-\Pmax}{blue!65!black}
  \ant{2}{\y}{1}{\rmin}{\Pmax}{blue!65!black}
  \ant{3}{\y}{1}{\rmin}{\Pmax}{blue!65!black}
  \ant{4}{\y}{-1}{\rmin}{-\Pmax}{blue!65!black}
  \node at (\colA,\y+0.1) {maximal}; \node[blue!65!black] at (\colB,\y+0.1) {$0$};

  \def\y{-4.2}
  \node[anchor=south west, font=\footnotesize\itshape] at (-0.35,\y+0.47) {(d) worst beamforming: split};
  \ant{1}{\y}{1}{\rmin}{\Pmax}{blue!65!black}
  \ant{2}{\y}{1}{\rmin}{\Pmax}{blue!65!black}
  \ant{3}{\y}{-1}{\rmin}{-\Pmax}{blue!65!black}
  \ant{4}{\y}{-1}{\rmin}{-\Pmax}{blue!65!black}
  \node at (\colA,\y+0.1) {maximal}; \node[blue!65!black] at (\colB,\y+0.1) {$\approx 0.25\,\Psi_{\max}$};

  \def\y{-5.6}
  \node[anchor=south west, font=\footnotesize\itshape] at (-0.35,\y+0.47) {(e) worst localization};
  \ant{1}{\y}{1}{\rmin}{\Pmax}{orange!85!black}
  \ant{2}{\y}{1}{\rmax}{\Pmax}{orange!85!black}
  \ant{3}{\y}{-1}{\rmax}{-\Pmax}{orange!85!black}
  \ant{4}{\y}{-1}{\rmin}{-\Pmax}{orange!85!black}
  \node at (\colA,\y+0.1) {smaller}; \node[orange!80!black] at (\colB,\y+0.1) {$\approx 0.33\,\Psi_{\max}$ (largest)};

  \begin{scope}[shift={(0,-6.4)}]
    \draw[black!50, densely dashed] (0,0) circle (\rmax);
    \node[anchor=west] at (0.2,0) {nominal position};
    \fill[black!70] (2.85,0) circle (\rmax);
    \node[anchor=west] at (3.02,0) {unit gain};
    \fill[black!70] (4.75,0) circle (\rmin);
    \node[anchor=west] at (4.88,0) {gain $1-\delta_g$};
    \fill[black!12] (0,-0.47) ++(-\dpd,-0.035) rectangle ++(2*\dpd,0.07);
    \draw[black!45] (-\dpd,-0.54) -- (-\dpd,-0.4) (\dpd,-0.54) -- (\dpd,-0.4);
    \node[anchor=west] at (0.38,-0.47) {range of position error};
    \draw[->, line width=0.7pt, black!70] (3.85,-0.62) -- ++(90-\Pmax:0.4);
    \node[anchor=west] at (4.1,-0.47) {extra phase $\Psi_k\in[-\Psi_{\max},\Psi_{\max}]$};
  \end{scope}
\end{tikzpicture}
\caption{Impairment configurations for $N=4$ ($\delta_g=0.7$, $\cos(\theta_0)>0$): (a) nominal array; (b) random impairments; (c)--(d) two worst cases for beamforming toward $\theta_0$ (Theorem~\ref{theo:bf}), with the same maximal loss at $\theta_0$ but very different AoA errors; (e) worst case for localization (Theorem~\ref{theo:loc}, $\varsigma=+1$). AoA errors are given in the natural variable $\Delta_0$.}
\label{fig:worst_ULA}
\end{figure}

Since the two functions are sensitive to the two components of the budget of Lemma~\ref{lem:budget}, their worst cases are nearly disjoint, as illustrated in Fig.~\ref{fig:worst_ULA} for $N=4$.
\begin{itemize}
    \item \emph{The worst case for beamforming toward the strongest direction does not shift the peak} (Theorem~\ref{theo:bf}(ii)). A calibration check based only on the AoA accuracy, e.g., on the peak position for a reference source at a known angle, would declare such an array perfectly calibrated, although its beamforming gain is minimal; with noise, it also degrades the AoA accuracy, as shown in Section~\ref{sec:noise}.
    \item \emph{The worst case for localization loses little beamforming gain.} It spends most of the budget on the tilt (for unit gains and large $N$, a phase step between two halves puts $75\%$ of the dispersion in the tilt, as in Section~\ref{sec:stat}), and keeps the central gains at their nominal value. For instance, with $N=8$, $\delta_p=\alpha_g=0.08$, $\theta_0=47^\circ$, and $\delta_g=0.7$, its gain toward the strongest direction is $-4.7$~dB ($-7.3$~dB toward $\theta_0$), against $-14.0$~dB for the worst case for beamforming. A calibration procedure that only monitors the beamforming gain may therefore miss the configurations that are the most harmful for localization, in line with the observation of \cite{10824218} that communication-oriented calibration may leave residual impairments affecting sensing.
    \item \emph{The two worst cases only meet through the split pattern, and only toward $\theta_0$.} Among the worst cases for beamforming toward the true angle, the split pattern maximizes the tilt; the worst case for localization keeps its phase arrangement and raises its central gains, which increases the peak shift by the lever factor $\Lambda(\delta_g)$.
\end{itemize}
Hence, neither the beamforming gain nor the AoA accuracy alone is a sufficient calibration criterion for a dual-function array.

\begin{remark}[Implications for calibration]
\label{rem:calibration}
Calibration procedures estimate the impairments by optimizing a cost function, and the decomposition tells which components each cost can see. A beamforming cost in which the beam direction is free, such as the received power after beam sweeping or the fit of measured channels by steering vectors at unknown angles, is exactly invariant to the tilt by \eqref{eq:gpeak}: any tilt is absorbed by a re-pointing of the beam, so that the tilt lies in the null space of such a cost. Conversely, a localization cost based on the estimated angle of a reference source only sees the tilt, the residual affecting the peak location at third order only (Proposition~\ref{prop:residual}(i)). Hence, a calibration based only on a beamforming cost with a free beam direction cannot correct the impairments that are the most harmful for localization, and a calibration based only on the AoA of reference sources cannot correct those that are the most harmful for beamforming. This is consistent with the well-known ambiguity of the blind calibration of ULAs with unknown gains and phases, for which the source directions can only be determined up to a common rotation \cite{astely1999spatial,cao2015spatial}: in the natural variable, this rotation is precisely the tilt. Position errors along the array axis add a second ambiguity, since a uniform stretch of the array yields a tilt proportional to $\cos(\theta_0)$, i.e., a dilation of all $\cos\theta$. Correcting the tilt thus requires an absolute reference, such as a source at a known direction or a measurement of the array geometry, whereas the residual can be calibrated blindly. Depending on their cost function and on whether the source positions are known, the learning-based calibration methods mentioned in Section~\ref{sec:intro} \cite{yassine2022mpnet,klaimi2026mompnet,chatelier2025diffmusic,10824218} may thus correct one component, the other, or both.
\end{remark}

The worst case for localization is moreover a plausible configuration in practice, for three reasons.
\begin{itemize}
    \item \emph{It only requires a single defect.} Its phase and position errors are constant over each half of the array: they are produced by one phase offset between two subarrays and one spacing error at their junction, not by $N$ independent errors reaching their bounds simultaneously. Large arrays are routinely assembled from panels or tiles, each with its own radio-frequency chains, oscillator distribution, calibration, and mechanical mounting \cite{9419749,6760595}, which creates exactly this kind of step.
    \item \emph{Weaker edge antennas are common.} In a finite array, the edge antennas do not see the same electromagnetic environment as the central ones, because of mutual coupling, and their gains typically differ \cite{9135245}. The gain profile of Theorem~\ref{theo:loc} precisely exhibits this behavior.
    \item \emph{Correlated errors do not average out}, as shown next.
\end{itemize}

\subsection{Typical Allocations: Independent and Panel-Level Errors}
\label{sec:stat}

Random impairments spend the budget of Lemma~\ref{lem:budget} at random. How much of it goes into the tilt depends on how the errors are correlated across the antennas. The following proposition considers $M$ contiguous panels of $\frac NM$ antennas each, with an extra phase that is constant within each panel and independent across panels; $M=N$ corresponds to errors that are independent across antennas.

\begin{proposition}[Tilt of random panel errors]
\label{prop:panels}
With unit gains, let the extra phases be constant within each of $M$ contiguous panels, and independent across panels, with zero mean and variance $\sigma_\Psi^2$. Then $\mathbb{E}\{\Delta_0^\star\}=0$, and
\begin{equation}
\label{eq:panel_var}
\mathbb{E}\{(\Delta_0^\star)^2\}=\frac{\sigma_\Psi^2\sum_{m=1}^MS_m^2}{(2\pi)^2\big(\sum_k(d_k-\bar d)^2\big)^2},
\end{equation}
with $S_m\triangleq\sum_{k\in\text{panel }m}(d_k-\bar d)$.
For large $N$, relative to the main-lobe half-width,
\begin{equation}
\label{eq:panel_rel}
\frac{\sqrt{\mathbb{E}\{(\Delta_0^\star)^2\}}}{2/N}\approx\frac{\sigma_\Psi}{\pi}\sqrt{\frac{3(M^2-1)}{M^3}},
\end{equation}
and the tilt captures, on average, a fraction $\frac{1}{N-1}$ of the dispersion of the phase errors for $M=N$, and approximately $\frac{M+1}{M^2}$ for fixed $M$.
\end{proposition}
\begin{proof}
See Appendix~\ref{app:panels}.
\end{proof}

For errors independent across antennas ($M=N$), the tilt only captures a fraction $\frac{1}{N-1}$ of the dispersion: almost all of it goes into the residual. The beamforming gain degrades, with a raised sidelobe floor, as described by antenna tolerance theory \cite{ruze1966antenna}, whereas the AoA error averages out: by \eqref{eq:panel_rel}, it decreases as $N^{-1/2}$ relative to the main-lobe half-width, i.e., as $N^{-3/2}$ in absolute terms. For uniform errors, $\sigma_\Psi^2=\frac{(2\pi)^2}{3}\big(\alpha_g^2+\delta_p^2\cos^2(\theta_0)\big)$. For panel-level errors, by contrast, \eqref{eq:panel_rel} does not depend on $N$: the error remains a fixed fraction of the main-lobe width, whatever the size of the array. With two panels, $75\%$ of the dispersion goes into the tilt; the worst case of Theorem~\ref{theo:loc} is the extreme case where both panels reach their bounds. In panel-based arrays, inter-panel calibration errors therefore dominate the localization accuracy.

On the other hand, the mean AoA error is zero, since replacing all $\Psi_k$ by $-\Psi_k$ leaves the distribution of the impairments unchanged and mirrors the ambiguity function about $\Delta_0=0$. For errors independent across antennas, the tilt is a sum of many independent terms, hence approximately Gaussian, so that the mean absolute error is $\sqrt{2/\pi}$ times the root mean square error of Proposition~\ref{prop:panels}. Gain errors and the nonlinearity of the phasors only change it by a few percent in the settings considered here.

\section{Effect of Noise}
\label{sec:noise}

With noise, the BS estimates the AoA from the noisy correlation $z(\theta)=\frac{\beta}{\sqrt N}v(\Delta_0)+\boldsymbol{w}(\theta)^H\boldsymbol{n}$, which yields a random estimate $\tilde\Delta_0$. Its error has three components, one for each consequence of the impairments: a bias, due to the peak shift (C1); a local variance around the biased value, due to the peak blurring (C2); and outliers, when the noise makes a sidelobe exceed the main lobe, whose occurrence is aggravated by the sidelobe rise (C3).

\subsection{Local Regime: Bias and Variance}
\label{sec:local}

At high SNR, the estimate stays in the main lobe, close to the noiseless peak $\hat\Delta_0$, which is its mean up to $O(1/\mathrm{SNR})$: the bias is the peak shift of Section~\ref{sec:effects}. Its variance follows from a linearization of the noisy peak around $\hat\Delta_0$.

\begin{proposition}[Local variance]
\label{prop:locvar}
At high SNR,
\begin{equation}
\label{eq:locvar}
\mathrm{Var}(\tilde\Delta_0)\approx\frac{1}{2\,\mathrm{SNR}}\,\frac{\sum_k\big|v'(\hat\Delta_0)-j2\pi d_k\,v(\hat\Delta_0)\big|^2}{\Big[\mathrm{Re}\big(|v'(\hat\Delta_0)|^2+v(\hat\Delta_0)^*v''(\hat\Delta_0)\big)\Big]^2}.
\end{equation}
For a nominal array, \eqref{eq:locvar} is the Cram\'er--Rao bound of single-tone frequency estimation \cite{rife1974single}, with the spatial frequency $\pi\Delta_0$: $\mathrm{CRB}\triangleq\frac{24}{(2\pi)^2N(N^2-1)\,\mathrm{SNR}}$. For small residuals,
\begin{equation}
\label{eq:locvar_approx}
\mathrm{Var}(\tilde\Delta_0)\approx\mathrm{CRB}\cdot\frac{\Gamma_\rho}{1-\mathrm{Var}_\rho(r)},
\end{equation}
where $\Gamma_\rho\triangleq\frac{\sum_k(d_k-\bar d)^2\,\sum_k(d_k-c)^2}{(\sum_k\rho_k(d_k-\bar d)^2)^2}\ge1$, with equality for unit gains.
\end{proposition}
\begin{proof}
See Appendix~\ref{app:locvar}.
\end{proof}

Beam scanning reaches this bound, since for a nominal array it is the maximum-likelihood estimator \cite{rife1974single}. The bound has a simple interpretation in terms of the tilt: at high SNR, the noise perturbs the phase of each antenna by an independent error of variance $\frac{1}{2\,\mathrm{SNR}}$, and the estimate is the least-squares slope of these phases across the aperture, i.e., a random tilt, whose variance $\frac{1}{2\,\mathrm{SNR}}\big/\big((2\pi)^2\sum_k(d_k-c)^2\big)$ is the CRB. It decreases as $N^3$ because $\sum_k(d_k-c)^2=\frac{N(N^2-1)}{48}$ grows both with the number of antennas and with the square of the aperture, which acts as the lever arm of the slope. This is also why independent hardware phase errors behave, for localization, like noise: by Proposition~\ref{prop:panels} with $M=N$, they produce a random tilt with the same variance as the CRB at the equivalent SNR $\frac{1}{2\sigma_\Psi^2}$, except that this tilt is frozen for a given array. The tilt does not appear in \eqref{eq:locvar_approx}: the peak shift moves the mean of the estimate, not its spread. The local variance is increased by the peak blurring only, through its two causes: the gains, through the factor $\Gamma_\rho$, which equals $1$ for unit gains and grows when the gains lower the peak and shrink the effective aperture; and the residual, through the factor $\frac{1}{1-\mathrm{Var}_\rho(r)}$, which is the loss of peak height of Proposition~\ref{prop:residual}(ii). We checked that \eqref{eq:locvar} matches Monte Carlo simulations within a few percent, and that \eqref{eq:locvar_approx} matches \eqref{eq:locvar} within $0.2$~dB for $\Psi_{\max}\le30^\circ$, including for the worst case for beamforming. In this regime, the mean squared error is
\begin{equation}
\label{eq:mse}
\mathbb{E}\{\tilde\Delta_0^2\}\approx\hat\Delta_0^2+\mathrm{Var}(\tilde\Delta_0),
\end{equation}
dominated by the bias (C1) at high SNR, and by the variance (C2) at moderate SNR.

\subsection{Global Regime: Outliers and Threshold}
\label{sec:threshold}

Below a threshold SNR, the noise sometimes lifts a sidelobe above the main lobe, and the estimate jumps to it: the error then departs from \eqref{eq:mse}. This threshold effect exists for any array, since even a nominal array has sidelobes, about $13$~dB below its peak; it has been analyzed for a nominal array in \cite{rife1974single}. The impairments raise the threshold in two ways: the peak blurring lowers the peak, hence the effective SNR at the peak, $N\,\mathrm{SNR}\,|v(\hat\Delta_0)|^2/N^2$; and the sidelobe rise reduces the margin between the peak and the highest sidelobe. The threshold has no simple closed form, its numerical behavior is illustrated in Section~\ref{sec:simu}. Since the strongest sidelobes are generally not symmetric about the peak, outliers also shift the mean of the estimate: with noise, the mean error is no longer the noiseless bias.

\subsection{Consequence for Communication}
\label{sec:noise_comm}

When the BS beamforms toward the estimated angle, its gain is $\tilde G\triangleq|v(\tilde\Delta_0)|^2/N^2$. At high SNR, the beam points at the noiseless peak, so that $\tilde G\to G_\mathrm{peak}$: the bias costs nothing to communication, since the beam follows the shifted peak. Above the threshold, the parabola of Proposition~\ref{prop:residual}(ii) and Proposition~\ref{prop:locvar} give the pointing loss
\begin{equation}
\label{eq:pointing}
\mathbb{E}\{\tilde G\}\approx G_\mathrm{peak}\Big(1-(2\pi)^2\,\mathrm{Var}_\rho(d)\,\mathrm{Var}(\tilde\Delta_0)\Big),
\end{equation}
which is the curvature of the main lobe times the variance of the estimate. For a nominal array, it reduces to $1-\frac{1}{2N\,\mathrm{SNR}}$: pointing at the estimated angle costs very little. Below the threshold, by contrast, the beam sometimes points at a sidelobe, and the loss becomes large: the outliers of the sensing step, and hence the sidelobe rise, propagate to communication. In short, noise turns the peak blurring and the sidelobe rise into accuracy losses for both functions, whereas the peak shift remains specific to localization.

\section{Numerical Results}
\label{sec:simu}

This section illustrates the consequences C1 to C3 and their effect on the two functions of the array. Unless stated otherwise, we consider $N=16$, $\Psi_{\max}=40^\circ$, and $\theta_0=47^\circ$, which only sets the visible region. The impairments are described directly by the extra phases $\Psi_k$, which combine phase and position errors, and AoA errors are given in the natural variable $\Delta_0$, in main-lobe half-widths $\frac2N$, which makes them independent of $\theta_0$. Four characteristic configurations are considered, which extend those of Fig.~\ref{fig:worst_ULA} to $N=16$: (a) the split pattern of Fig.~\ref{fig:worst_ULA}(d), i.e., a phase step between the two halves of the array, $\Psi_k=\Psi_{\max}\mathrm{sgn}(d_k-c)$, with unit gains, which is also the worst case for localization for unit gains; (b) random independent errors, as in Fig.~\ref{fig:worst_ULA}(b), with $\Psi_k$ uniform in $[-\Psi_{\max},\Psi_{\max}]$ and $\rho_k$ uniform in $[0.7,1]$; (c) four panels with alternating offsets $\pm\Psi_{\max}$, with unit gains, which has no counterpart for $N=4$; and (d) the symmetric worst case for beamforming of Fig.~\ref{fig:worst_ULA}(c), with the pattern $(-,+,+,-)$ repeated and gains $0.7$. The peak of the correlation is searched over the whole visible region, i.e., one period of $\Delta_0$, on a grid of $4096$ values of $\Delta_0$ uniformly spaced in $\cos\theta$, computed by a zero-padded fast Fourier transform, and then refined by a golden-section search on the exact correlation. With noise, the curves are averaged over $2000$ noise draws.

\subsection{The Three Consequences}
\label{sec:simu_effects}

\begin{figure}[t]
\centering
\includegraphics[width=\columnwidth]{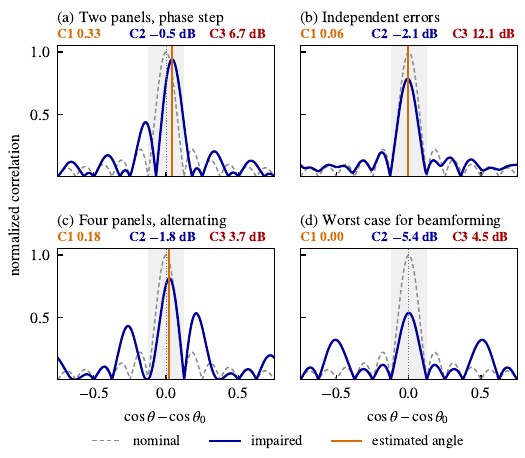}
\caption{Ambiguity function of the four configurations ($N=16$, $\Psi_{\max}=40^\circ$): C1 is the peak shift in main-lobe half-widths, C2 the peak height, and C3 the peak-to-sidelobe ratio ($13.1$~dB for the nominal array).}
\label{fig:simu_effects}
\end{figure}

Fig.~\ref{fig:simu_effects} shows the ambiguity function of the four configurations. Each configuration mostly exhibits one consequence, in line with Section~\ref{sec:consequences}. The phase step between two panels (a) puts $75\%$ of the dispersion in the tilt: it shifts the peak by a third of a half-width, but only lowers it by $0.5$~dB. Its residual is not zero, however, and raises a sidelobe on one side of the peak, through the interference term of \eqref{eq:mod2}. Independent errors (b) put most of the dispersion in the residual, with a white spectrum: they lower the peak by $2.1$~dB, barely shift it, and raise a uniform floor of sidelobes. Panels with alternating offsets (c) create a periodic residual, which concentrates its energy in two spurious lobes, only $3.7$~dB below the peak. Finally, the worst case for beamforming (d) spends the whole budget on the residual: it does not shift the peak at all, but lowers it by $5.4$~dB and creates two replicas at $\Delta_0=\pm\frac12$.

\subsection{Two Nearly Disjoint Worst Cases}
\label{sec:simu_disjoint}

\begin{figure}[t]
\centering
\includegraphics[width=\columnwidth]{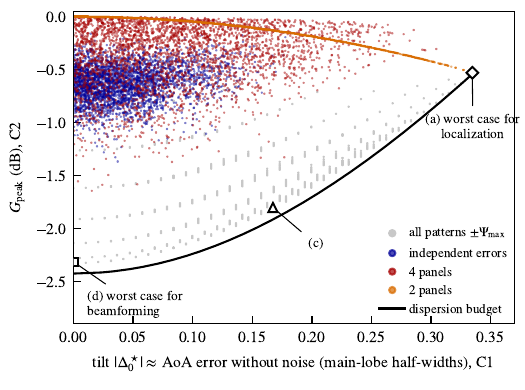}
\caption{Gain toward the strongest direction versus tilt, i.e., noiseless AoA error up to third order in the residual, for unit gains ($N=16$, $\Psi_{\max}=40^\circ$): all $2^N$ sign patterns $\pm\Psi_{\max}$, $3000$ random configurations of each type, and configurations (a), (c), (d) of Fig.~\ref{fig:simu_effects}. The dispersion budget forbids the region below the black curve.}
\label{fig:simu_disjoint}
\end{figure}

Fig.~\ref{fig:simu_disjoint} places impairment configurations in the plane formed by the AoA error without noise (C1) and the gain toward the strongest direction (C2), for unit gains. The horizontal axis is the tilt, which equals the AoA error up to third order in the residual (Proposition~\ref{prop:residual}(i)). The black curve is the dispersion budget: for unit gains, $|v_\mathrm{r}(0)|\ge\sum_k\cos r_k\ge N\big(1-\frac12\mathrm{Var}(r)\big)$, so that $G_\mathrm{peak}\ge\big(1-\frac12\mathrm{Var}(r)\big)^2$ by \eqref{eq:gpeak}, and Lemma~\ref{lem:budget} gives $\mathrm{Var}(r)\le\Psi_{\max}^2-(2\pi\Delta_0^\star)^2\mathrm{Var}(d)$. No configuration can lie below this curve, and the sign patterns at the bounds closely follow it. Configurations (a), (c), and (d) of Fig.~\ref{fig:simu_effects} are marked, with unit gains; configuration (b), which has gain errors, belongs to the cloud of independent errors. The two worst cases lie at the two ends of the curve: the worst case for beamforming (d) has no AoA error and the lowest gain, $-2.3$~dB, whereas the worst case for localization (a) has the largest AoA error, $0.33$ half-width, and loses only $0.5$~dB. The panels with alternating offsets (c) lie in between. No configuration is bad for both functions at the same time. Random configurations illustrate how the budget is typically spent: independent errors stay close to the vertical axis, since almost all their dispersion goes into the residual, whereas phase errors between two panels follow a curve that leads to the worst case for localization.

\subsection{Effect of Noise on Localization}
\label{sec:simu_aoa}

\begin{figure}[t]
\centering
\includegraphics[width=\columnwidth]{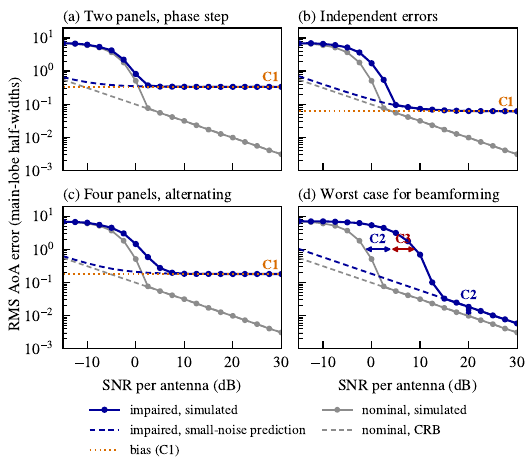}
\caption{RMS AoA error versus SNR for the four configurations of Fig.~\ref{fig:simu_effects}, with the small-noise prediction $\sqrt{\hat\Delta_0^2+\mathrm{Var}(\tilde\Delta_0)}$ of \eqref{eq:mse} and Proposition~\ref{prop:locvar}. The horizontal arrows show the shift of the threshold at an error of one half-width, split into the part explained by the peak loss (C2) and the remainder (C3).}
\label{fig:simu_aoa}
\end{figure}

Fig.~\ref{fig:simu_aoa} shows the RMS AoA error versus SNR. Above the threshold, all curves follow the small-noise prediction, and the nominal array reaches the CRB, which validates Proposition~\ref{prop:locvar}. Each consequence leaves its own mark on the curves. The peak shift sets a floor, the bias, which dominates at high SNR for the configurations (a) to (c): for the phase step between two panels, the error reaches this floor from about $2$~dB, so that the impairments dominate the accuracy at any practical SNR. The peak blurring increases the local variance: for the worst case for beamforming (d), which has no bias, the error remains about twice the CRB at high SNR, in agreement with \eqref{eq:locvar_approx}, where $\Gamma_\rho\approx2$ for gains $0.7$ and $\frac{1}{1-\mathrm{Var}_\rho(r)}\approx2$. Finally, the threshold moves to higher SNR. For (b) and (c), it moves by about $2.5$~dB, mostly explained by the peak loss. For (d), it moves by about $10$~dB, of which $5.4$~dB are explained by the peak loss and the rest by the sidelobe rise: although this configuration has no bias, it makes the AoA estimate unreliable below about $9$~dB per antenna, instead of $-1$~dB for the nominal array.

\subsection{Effect of Noise on Communication}
\label{sec:simu_comm}

\begin{figure}[t]
\centering
\includegraphics[width=\columnwidth]{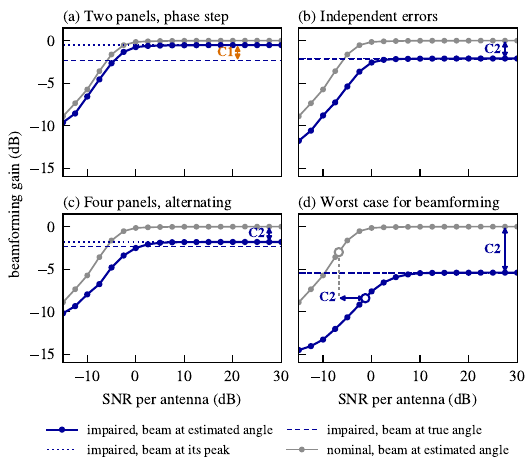}
\caption{Beamforming gain versus SNR when the beam points at the estimated angle, for the four configurations of Fig.~\ref{fig:simu_effects}, relative to a nominal array perfectly pointed at the UE. The horizontal arrow in (d) shows the shift of the threshold between the points (circles) where the nominal and impaired curves are $3$~dB below their high-SNR limits.}
\label{fig:simu_comm}
\end{figure}

Fig.~\ref{fig:simu_comm} shows the beamforming gain when the BS points its beam at the estimated angle. At high SNR, the gain reaches $G_\mathrm{peak}$, which only depends on the peak blurring. The bias costs nothing to communication: for the phase step between two panels (a), pointing at the estimated angle loses $0.5$~dB, whereas pointing at the true angle would lose $2.3$~dB, since the beam follows the shifted peak. At low SNR, the errors of the sensing step propagate to communication, and the gain drops below that of the nominal array, down to $-14.5$~dB at $-15$~dB per antenna for the worst case for beamforming. Its threshold moves by about $5.5$~dB, which is entirely explained by the peak loss: communication is much less sensitive than localization to the sidelobe rise, since a beam pointed at a nearby sidelobe loses less gain than the estimate loses accuracy.

\subsection{Independent Versus Panel-Level Errors}
\label{sec:simu_panels}

\begin{figure}[t]
\centering
\includegraphics[width=\columnwidth]{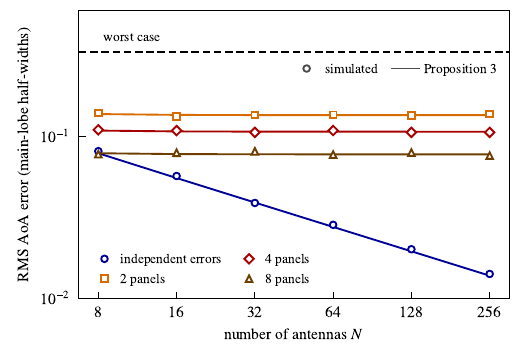}
\caption{RMS AoA error without noise versus $N$, for random errors independent across antennas or common to the antennas of each of $M$ panels (unit gains, $\Psi_{\max}=40^\circ$, $2000$ draws), with the prediction of Proposition~\ref{prop:panels} and the worst case of Theorem~\ref{theo:locerr}.}
\label{fig:simu_panels}
\end{figure}

Fig.~\ref{fig:simu_panels} shows the RMS AoA error without noise versus the number of antennas, for random errors independent across antennas or common to the antennas of each panel. The simulations match Proposition~\ref{prop:panels} within a few percent. Independent errors average out: their relative error decreases as $N^{-1/2}$, from $0.08$ half-width for $N=8$ to $0.014$ for $N=256$. Panel-level errors do not: their relative error only depends on the number of panels, about $0.14$, $0.11$, and $0.08$ half-width for $2$, $4$, and $8$ panels, whatever $N$, and the worst case does not depend on $N$ either, as predicted by Theorem~\ref{theo:locerr}. With $256$ antennas assembled in two panels, the error is thus about ten times larger than with independent errors of the same amplitude: in large modular arrays, inter-panel calibration matters much more than the calibration of individual antennas.

\section{Conclusion}
\label{sec:conclusion}

In this paper, we studied how gain, phase, and position errors affect the two functions of an ISAC array. Decomposing the phase errors into a tilt and a residual explains the three consequences of the impairments: the tilt only re-points the array (C1), whereas the residual lowers and widens the peak (C2) and puts the energy it takes from it in the sidelobes (C3); the gains act as a lever on these effects. Beamforming toward the strongest direction is exactly blind to the tilt, whereas localization only sees the tilt without noise, and also suffers from C2 and C3 with noise. Since the hardware bounds fix a budget on the dispersion of the phase errors, the worst cases for the two functions are nearly disjoint. The worst-case AoA error is a fixed fraction of the main-lobe width whatever the number of antennas, and errors common to the antennas of a panel do not average out, so that inter-panel calibration matters most in large modular arrays.

These results have a direct implication for calibration: a calibration based on a single function structurally leaves the impairments that are the most harmful for the other. Correcting the tilt requires an absolute reference, such as sources at known directions, whereas the residual can be calibrated blindly. Designing and evaluating calibration procedures that combine both is left for future work, as are monostatic sensing, where the impairments affect both transmission and reception, planar arrays, and near-field and multi-source scenarios.

\appendices

\section{Proof of Lemma~\ref{lem:energy}}
\label{app:energy}

By \eqref{eq:v}, $|v(\Delta_0)|^2=\sum_{k,l}\rho_k\rho_le^{-j(\Psi_k-\Psi_l)}e^{j\pi(k-l)\Delta_0}$. For any $a$, $\int_a^{a+2}e^{j\pi(k-l)\Delta_0}d\Delta_0$ equals $2$ if $k=l$ and $0$ otherwise. Integrating over the visible region, which has length $2$, thus only keeps the terms $k=l$, which gives \eqref{eq:energy}.

\section{Proof of Proposition~\ref{prop:residual}}
\label{app:residual}

With $x=\Delta_0-\Delta_0^\star$ as in Proposition~\ref{prop:residual}, Proposition~\ref{prop:factorization} gives $|v(\Delta_0)|=|v_\mathrm{r}(x)|$.

\emph{Second-order pattern.} For any real $r$, $|e^{-jr}-1+jr+\frac{r^2}{2}|\le\frac{|r|^3}{6}$. Hence, for all $x$,
\begin{align}
\left|v_\mathrm{r}(x)-v_2(x)\right|&=\Big|\sum_k\rho_ke^{j2\pi d_kx}\big(e^{-jr_k}-1+jr_k+\tfrac{r_k^2}{2}\big)\Big|\nonumber\\
&\le\frac16\sum_k\rho_k|r_k|^3\le\frac{r_{\max}}{6}\sum_k\rho_kr_k^2,
\end{align}
and $\sum_k\rho_kr_k^2=P\,\mathrm{Var}_\rho(r)$ since the weighted mean of the residual is zero, which gives \eqref{eq:v2bound} by Proposition~\ref{prop:factorization}. Expanding $|v_2|^2=|V_\rho-jR-\frac12R_2|^2$ and dropping the terms of order three and more in the residual gives \eqref{eq:mod2}.

\emph{(i) Peak location.} At $x=0$, $V_\rho(0)=P$ and $R_2(0)=\sum_k\rho_kr_k^2$ are real, whereas $V_\rho'(0)=j2\pi P\bar d$ and $R_2'(0)=j2\pi\sum_k\rho_kd_kr_k^2$ are purely imaginary. Hence the slopes of $|V_\rho|^2$, i.e., $2\,\mathrm{Re}\big(V_\rho(0)^*V_\rho'(0)\big)$, and of $\mathrm{Re}(V_\rho^*R_2)$, i.e., $\mathrm{Re}\big(V_\rho'(0)^*R_2(0)+V_\rho(0)^*R_2'(0)\big)$, are zero, and so are those of $|R|^2$ and $\mathrm{Im}(V_\rho^*R)$, since $R(0)=R'(0)=0$. To find the third-order shift, let $F(x)\triangleq|v_\mathrm{r}(x)|^2=\sum_{k,l}\rho_k\rho_l\cos\big(2\pi(d_k-d_l)x-(r_k-r_l)\big)$, whose derivatives at $x=0$ are
\begin{align}
F'(0)&=2\pi\sum_{k,l}\rho_k\rho_l(d_k-d_l)\sin(r_k-r_l),\nonumber\\
F''(0)&=-(2\pi)^2\sum_{k,l}\rho_k\rho_l(d_k-d_l)^2\cos(r_k-r_l).\nonumber
\end{align}
Since $\sum_{k,l}\rho_k\rho_l(d_k-d_l)(r_k-r_l)=2P\sum_k\rho_k(d_k-\bar d)r_k=0$ by orthogonality, and $\sin u=u-\frac{u^3}{6}+O(u^5)$, $F'(0)=-\frac{\pi}{3}\sum_{k,l}\rho_k\rho_l(d_k-d_l)(r_k-r_l)^3+O(r_{\max}^5)$, whereas $F''(0)=-(2\pi)^2\sum_{k,l}\rho_k\rho_l(d_k-d_l)^2+O(r_{\max}^2)$ is negative and bounded away from zero. The maximum of $F$ in the main lobe is therefore located at $x=-F'(0)/F''(0)+O(F'(0)^2)$, which gives \eqref{eq:cubic}. For a pure tilt, Corollary~\ref{cor:tilt} applies; for symmetric patterns, see below.

\emph{(ii) Main lobe.} By Proposition~\ref{prop:factorization}, $|v(\Delta_0)|^2=|v_\mathrm{r}(x)|^2=\sum_{k,l}\rho_k\rho_l\cos(\phi_k-\phi_l)$. Since $\frac12\sum_{k,l}\rho_k\rho_l(a_k-a_l)^2=P^2\,\mathrm{Var}_\rho(a)$ for any sequence $(a_k)$,
\begin{multline}
|v(\Delta_0)|^2-P^2\big(1-\mathrm{Var}_\rho(\phi)\big)\\
=\sum_{k,l}\rho_k\rho_l\Big[\cos(\phi_k-\phi_l)-1+\tfrac12(\phi_k-\phi_l)^2\Big].
\end{multline}
For any real $u$, $0\le\cos u-1+\frac{u^2}{2}\le\frac{u^4}{24}$, so that this difference lies between $0$ and $P^2\xi$. Finally, the residual being orthogonal to $d_k-\bar d$, $\mathrm{Var}_\rho(\phi)=(2\pi)^2x^2\,\mathrm{Var}_\rho(d)+\mathrm{Var}_\rho(r)$, which gives \eqref{eq:parabola}. The same argument, applied to $|v(0)|^2=\sum_{k,l}\rho_k\rho_l\cos(\Psi_k-\Psi_l)$, gives \eqref{eq:v0}.

\emph{(iii) Energy balance.} Over any interval of length $2$, the mean of $e^{j2\pi(d_k-d_l)x}$ is $1$ if $k=l$ and $0$ otherwise. Hence the mean of $V_\rho^*R_2$ is $\sum_k\rho_k\cdot\rho_kr_k^2$, which is real, the mean of $V_\rho^*R$ is $\sum_k\rho_k^2r_k$, also real, so that the mean of its imaginary part is zero, and the means of $|V_\rho|^2$ and $|R|^2$ are $\sum_k\rho_k^2$ and $\sum_k\rho_k^2r_k^2$.

\emph{Symmetric patterns.} If $\rho_{N+1-k}=\rho_k$ and $\Psi_{N+1-k}=\Psi_k$ for all $k$, let $c\triangleq\frac{N-1}{4}$ and $e_k\triangleq d_k-c$, so that $e_{N+1-k}=-e_k$. Pairing the antennas $k$ and $N+1-k$ in \eqref{eq:v} gives $v(\Delta_0)=e^{j2\pi c\Delta_0}\sum_k\rho_ke^{-j\Psi_k}\cos(2\pi e_k\Delta_0)$, so that $|v|$ is even, and $\Delta_0^\star=0$ since $\bar d=c$ and $\sum_k\rho_k(d_k-c)\Psi_k=0$.

\section{Proof of Theorem~\ref{theo:bf}}
\label{app:bf}

\emph{(i)} Since $\rho_k\ge1-\delta_g$ and $\cos\Psi_k\ge\cos\Psi_{\max}>0$, $|v(0)|\ge\mathrm{Re}\,v(0)=\sum_k\rho_k\cos\Psi_k\ge N(1-\delta_g)\cos\Psi_{\max}$. Equality requires $\rho_k=1-\delta_g$ and $|\Psi_k|=\Psi_{\max}$ for all $k$, and $\mathrm{Im}\,v(0)=-(1-\delta_g)\sin\Psi_{\max}\sum_kp_k=0$, i.e., balanced signs. Conversely, these conditions give $v(0)=N(1-\delta_g)\cos\Psi_{\max}$.

\emph{(ii)} $G_\mathrm{peak}\ge|v(0)|^2/N^2=G_{\theta_0}$, with equality if and only if the maximum of $|v|$ is reached at $\Delta_0=0$. Combining with (i) gives the result.

\section{Proof of Theorem~\ref{theo:loc}}
\label{app:loc}

By \eqref{eq:tilt_bound}, for any gain profile, the largest tilt is $\frac{\Psi_{\max}}{2\pi}Q(\boldsymbol\rho)$, reached by $\Psi_k=-\varsigma\Psi_{\max}\,\mathrm{sgn}(d_k-\bar d)$, $\varsigma\in\{-1,+1\}$, which gives the first two conditions of Theorem~\ref{theo:loc} since $\Psi_k=2\pi(\varepsilon_k\cos(\theta_0)-\varphi_k)$, and $\Delta_0^\star=-\varsigma\frac{\Psi_{\max}}{2\pi}Q(\boldsymbol\rho)$. It remains to maximize $Q$ over the gains.

\emph{Symmetric gain profiles.} For such profiles, $\bar d=c$ does not depend on the gains, and $Q(\boldsymbol\rho)=A(\boldsymbol\rho)/B(\boldsymbol\rho)$, with $A(\boldsymbol\rho)\triangleq\sum_k\rho_k|d_k-c|$ and $B(\boldsymbol\rho)\triangleq\sum_k\rho_k(d_k-c)^2$, is a ratio of linear functions of the gains. Its derivative with respect to $\rho_{k'}$ is
\begin{equation}
\frac{\partial Q}{\partial\rho_{k'}}=\frac{|e_{k'}|\sum_{k\neq k'}\rho_ke_k^2-e_{k'}^2\sum_{k\neq k'}\rho_k|e_k|}{B(\boldsymbol\rho)^2},
\end{equation}
with $e_k\triangleq d_k-c$, whose numerator does not depend on $\rho_{k'}$. $Q$ is thus monotonic in each gain, and its maximum is reached at a vertex of the box $[1-\delta_g,1]^N$. Moreover, $\mathrm{sgn}\big(\frac{\partial Q}{\partial\rho_k}\big)=\mathrm{sgn}\big(|d_k-c|B-(d_k-c)^2A\big)=\mathrm{sgn}\big(\frac{1}{|d_k-c|}-Q(\boldsymbol\rho)\big)$. At the maximum, $\rho_k=1$ when $|d_k-c|<\frac1{Q^\star}$, and $\rho_k=1-\delta_g$ when $|d_k-c|>\frac1{Q^\star}$.

\emph{Arbitrary gain profiles.} For any gain profile and any reference point $c'$, let $Q(\boldsymbol\rho,c')\triangleq\sum_k\rho_k|d_k-c'|/\sum_k\rho_k(d_k-c')^2$, so that $Q(\boldsymbol\rho)=Q(\boldsymbol\rho,\bar d)$. It suffices to show that $Q(\boldsymbol\rho,c')\le\gamma$ for all $\boldsymbol\rho$ and $c'$, where $\gamma$ denotes $Q^\star$ in the large-$N$ regime, i.e., the maximum of $Q(\cdot,c)$. Let $f(u)\triangleq|u|-\gamma u^2$. Since $Q(\boldsymbol\rho,c')\le\gamma$ if and only if $\sum_k\rho_kf(d_k-c')\le0$, we have to show that $W(c')\triangleq\max_{\boldsymbol\rho}\sum_k\rho_kf(d_k-c')\le0$. The maximum is reached with $\rho_k=1$ where $f>0$ and $\rho_k=1-\delta_g$ elsewhere, so that, with the Riemann approximation of Appendix~\ref{app:Qstar}, $W(c')\approx2\left[I(a)+I(\ell-a)\right]$, where $\ell\approx\frac N2$ is the aperture, $a$ the distance from $c'$ to the first antenna, and $I(t)\triangleq\int_0^t\omega(u)\,du$, with $\omega(u)=f(u)$ if $f(u)>0$ and $\omega(u)=(1-\delta_g)f(u)$ otherwise. At the center ($a=\frac\ell2$), $W=0$ by definition of $\gamma$. Moreover, $\frac{d}{da}\left[I(a)+I(\ell-a)\right]=\omega(a)-\omega(\ell-a)\ge0$ for $a<\frac\ell2$. Indeed, $\omega$ increases on $[0,\frac{1}{2\gamma}]$, decreases beyond, and is negative beyond $\frac1\gamma$; since $\gamma\ge\frac3\ell$ (the value obtained with uniform gains), $\ell-a>\frac\ell2>\frac1\gamma$, so that $\omega(\ell-a)<0\le\omega(a)$ if $a\le\frac1\gamma$, while $\omega(\ell-a)\le\omega(a)$ if $a>\frac1\gamma$, both points lying in the decreasing region. By symmetry, $I(a)+I(\ell-a)$ is thus maximal at $a=\frac\ell2$, so that $W(c')\le0$ for all $c'$: no gain profile yields a larger tilt than the configuration of Theorem~\ref{theo:loc}.

\section{Proof of Corollary~\ref{cor:Qstar} and Theorem~\ref{theo:locerr}}
\label{app:Qstar}

For a symmetric profile, the antennas $k>\frac N2$ are at distances $m_i\triangleq d_k-c=\frac{2i-1}{4}$ from the center, with $i=k-\frac N2\in\{1,\ldots,\frac N2\}$, so that $Q(\boldsymbol\rho)=\sum_i\rho(m_i)m_i/\sum_i\rho(m_i)m_i^2$. With the step $\frac12$ between consecutive $m_i$ and the half-aperture $m_{\max}\triangleq\frac N4$, both sums are approximated by Riemann integrals. With the threshold gain profile of Theorem~\ref{theo:loc},
\begin{align}
\sum_i\rho(m_i)m_i&\approx2\left(\int_0^{\frac1Q}m\,dm+(1-\delta_g)\int_{\frac1Q}^{m_{\max}}m\,dm\right)\nonumber\\
&=(1-\delta_g)m_{\max}^2+\frac{\delta_g}{Q^2},\\
\sum_i\rho(m_i)m_i^2&\approx\frac23\left((1-\delta_g)m_{\max}^3+\frac{\delta_g}{Q^3}\right).
\end{align}
$Q^\star$ is thus the fixed point of the ratio of these two expressions. With $\tau\triangleq\frac{1}{m_{\max}Q^\star}$, simple rearrangements give $\delta_g\tau^3+3(1-\delta_g)\tau-2(1-\delta_g)=0$. With $p\triangleq\frac{3(1-\delta_g)}{\delta_g}$ and $q\triangleq-\frac{2(1-\delta_g)}{\delta_g}$, the discriminant $D_c\triangleq(\frac q2)^2+(\frac p3)^3=\frac{(1-\delta_g)^2}{\delta_g^3}$ is positive, so that this cubic has a unique real root, $\tau=\sqrt[3]{-\frac q2+\sqrt{D_c}}+\sqrt[3]{-\frac q2-\sqrt{D_c}}$, which gives \eqref{eq:tau}. Then $Q^\star=\frac{1}{m_{\max}\tau}=\frac{4}{N\tau}$, and $M_\mathrm{uni}$, the number of antennas in $[0,\frac{1}{Q^\star}]$, is $\frac{2}{Q^\star}=\frac{\tau N}{2}$.

Finally, $\Delta_0^\mathrm{wst}=\frac{\Psi_{\max}}{2\pi}Q^\star\approx\frac{2\Psi_{\max}}{\pi N\tau}=\frac{3\Psi_{\max}}{\pi N}\cdot\frac{2}{3\tau}$, which is \eqref{eq:wst}, and $\Delta_0=\cos(\theta)-\cos(\theta_0)$ gives the angular expression.

\section{Proof of Proposition~\ref{prop:panels}}
\label{app:panels}

With unit gains, \eqref{eq:tilt} reads $\Delta_0^\star=\frac{\sum_k(d_k-\bar d)\Psi_k}{2\pi\sum_k(d_k-\bar d)^2}=\frac{\sum_mS_m\Psi^{(m)}}{2\pi\sum_k(d_k-\bar d)^2}$, where $\Psi^{(m)}$ is the extra phase of panel $m$. Since the $\Psi^{(m)}$ are independent with zero mean and variance $\sigma_\Psi^2$, $\mathbb{E}\{\Delta_0^\star\}=0$ and \eqref{eq:panel_var} follows. For $M=N$, $\sum_mS_m^2=\sum_k(d_k-\bar d)^2=\frac{N(N^2-1)}{48}$. For fixed $M$ and large $N$, $S_m\approx\frac{N^2}{2M^2}\big(m-\frac{M+1}{2}\big)$, so that $\sum_mS_m^2\approx\frac{N^4(M^2-1)}{48M^3}$, whereas $\sum_k(d_k-\bar d)^2\approx\frac{N^3}{48}$; this gives \eqref{eq:panel_rel}, which also holds for $M=N$ in the large-$N$ limit. Finally, the expected tilt part of the dispersion is $\mathbb{E}\{(2\pi\Delta_0^\star)^2\}\mathrm{Var}(d)=\frac{\sigma_\Psi^2\sum_mS_m^2}{N\sum_k(d_k-\bar d)^2}$, and the expected dispersion is $\mathbb{E}\{\mathrm{Var}(\Psi)\}=\sigma_\Psi^2(1-\frac1M)$. Their ratio is $\frac{1}{N-1}$ for $M=N$, and approximately $\frac{M+1}{M^2}$ for fixed $M$.

\section{Proof of Proposition~\ref{prop:locvar}}
\label{app:locvar}

With $s=1$, the noisy correlation is proportional to $v(\Delta_0)+\nu(\Delta_0)$, with $\nu(\Delta_0)\triangleq\sum_k\tilde n_ke^{j2\pi d_k\Delta_0}$, where the $\tilde n_k$ are independent $\mathcal{CN}(0,\frac{1}{\mathrm{SNR}})$, since multiplying the noise by known phases does not change its distribution. The estimate satisfies $\frac{d}{d\Delta_0}|v+\nu|^2=0$, i.e., $\mathrm{Re}\big((v+\nu)^*(v'+\nu')\big)=0$. The noiseless term vanishes at $\hat\Delta_0$; linearizing around $\hat\Delta_0$ to first order in the noise gives $\tilde\Delta_0-\hat\Delta_0\approx-\mathrm{Re}(v^*\nu'+v'^*\nu)/\mathrm{Re}(|v'|^2+v^*v'')$, evaluated at $\hat\Delta_0$. The numerator reads $\mathrm{Re}\big(\sum_k\tilde n_kc_k\big)$, with $c_k\triangleq e^{j2\pi d_k\hat\Delta_0}\big(v'^*+j2\pi d_kv^*\big)$, whose variance is $\frac{1}{2\,\mathrm{SNR}}\sum_k|c_k|^2=\frac{1}{2\,\mathrm{SNR}}\sum_k|v'-j2\pi d_kv|^2$, which gives \eqref{eq:locvar}. For a nominal array, $\hat\Delta_0=0$, $v(0)=N$, $v'(0)=j2\pi Nc$, and $\mathrm{Re}(|v'|^2+v^*v'')=-(2\pi)^2N\sum_k(d_k-c)^2$, so that \eqref{eq:locvar} becomes $\frac{1}{2(2\pi)^2\mathrm{SNR}\sum_k(d_k-c)^2}=\mathrm{CRB}$, with $\sum_k(d_k-c)^2=\frac{N(N^2-1)}{48}$.

For small residuals, by Proposition~\ref{prop:factorization}, $v'-j2\pi d_kv=j2\pi e^{j\chi}\big(B-d_kA\big)$ at the peak, with $A\triangleq\sum_l\rho_le^{-jr_l}$ and $B\triangleq\sum_l\rho_ld_le^{-jr_l}$. By orthogonality of the residual, $B/A=\bar d+O(r_{\max}^2)$, so that the numerator is approximately $(2\pi)^2|A|^2\sum_k(d_k-\bar d)^2$, with $|A|^2\approx P^2\big(1-\mathrm{Var}_\rho(r)\big)$ by Proposition~\ref{prop:residual}(ii). The denominator is half the second derivative of $|v|^2$ at its peak, $-(2\pi)^2\sum_{k,l}\rho_k\rho_l\frac{(d_k-d_l)^2}{2}\cos(r_k-r_l)$, which is approximately $-(2\pi)^2P^2\,\mathrm{Var}_\rho(d)\big(1-\mathrm{Var}_\rho(r)\big)$ when the residual is not correlated with the squared distances between antennas. Combining both gives $\mathrm{Var}(\tilde\Delta_0)\approx\frac{\sum_k(d_k-\bar d)^2}{2(2\pi)^2\mathrm{SNR}\,P^2\mathrm{Var}_\rho(d)^2(1-\mathrm{Var}_\rho(r))}$, which is \eqref{eq:locvar_approx}, since $P\,\mathrm{Var}_\rho(d)=\sum_k\rho_k(d_k-\bar d)^2$. Finally, $\Gamma_\rho\ge1$, since $\sum_k\rho_k(d_k-\bar d)^2$ is at most $\sum_k\rho_k(d_k-c)^2\le\sum_k(d_k-c)^2$, $\bar d$ minimizing the weighted sum and the gains being at most $1$, and at most $\sum_k(d_k-\bar d)^2$, for the same reason.

\bibliographystyle{IEEEtran}
\bibliography{mabiblio}
\end{document}